\documentclass[AMA,STIX1COL]{WileyNJD-v2}

\articletype{Original Article}%

\received{23 August 2019}
\revised{DD MMMM YYYY}
\accepted{DD MMMM YYYY}

\newcommand{\dataset}{{\cal D}}
\newcommand{\iid}{{\overset{\text{i.i.d.}}{\sim}}}

\newcommand\independent{\protect\mathpalette{\protect\independenT}{\perp}}
\def\independenT#1#2{\mathrel{\rlap{$#1#2$}\mkern2mu{#1#2}}}

\begin{document}

\title{A Pilot Design for Observational Studies: \\Using Abundant Data Thoughtfully}

\author[1,2]{Rachael C. Aikens}
\author[2]{Dylan Greaves}
\author[1,2,3]{Michael Baiocchi PhD}

\authormark{Aikens \textsc{et al}}

\address[1]{\orgdiv{Program in Biomedical Informatics}, \orgname{Stanford University}, \orgaddress{\state{California}, \country{USA}}}

\address[2]{\orgdiv{Department of Statistics}, \orgname{Stanford University}, \orgaddress{\state{California}, \country{USA}}}

\address[3]{\orgdiv{Department of Epidemiology and Population Health}, \orgname{Stanford University}, \orgaddress{\state{California}, \country{USA}}}

\corres{Michael Baiocchi, Redwood Building, 150 Governor's Lane, Stanford, CA 94305-5405 \email{baiocchi@stanford.edu}}

\presentaddress{Redwood Building, 150 Governor's Lane, Stanford, CA 94305-5405}

\abstract[Summary]{Observational studies often benefit from an abundance of observational units. This can lead to studies that -- while challenged by issues of internal validity -- have inferences derived from sample sizes substantially larger than randomized controlled trials. But is the information provided by an observational unit best used in the analysis phase? We propose the use of `pilot design,' in which observations are expended in the design phase of the study, and the post-treatment information from these observations is used to improve study design. In modern observational studies, which are data rich but control poor, pilot designs can be used to gain information about the structure of post-treatment variation. This information can then be used to improve instrumental variable designs, propensity score matching, doubly-robust estimation, and other observational study designs.  

We illustrate one version of a pilot design, which aims to reduce within-set heterogeneity and improve performance in sensitivity analyses. This version of a pilot design expends observational units during the design phase to fit a prognostic model, avoiding concerns of overfitting. Additionally, it enables the construction of `Assignment-Control (AC) plots,' which visualize the relationship between propensity and prognostic scores. We first show some examples of these plots, then we demonstrate in a simulation setting how this alternative use of the observations can lead to gains in terms of both treatment effect estimation and sensitivity analyses of unobserved confounding.}


\jnlcitation{\cname{
\author{R. Aikens}, 
\author{D. Greaves}, and
\author{M. Baiocchi}} (\cyear{2019}), 
\ctitle{A Pilot Design for Observational Studies: Using Abundant Data Thoughtfully}, \cjournal{}, \cvol{2019;XX:Y--Z}.}

\maketitle

\footnotetext{\textbf{Abbreviations:} SATE, sample average treatment effect; SATT, sample average treatment effect among the treated; MSE, mean squared error.}

\section{Introduction}\label{introduction}
Modern observational studies offer greater sample sizes than ever before. However, this often comes at the cost of great unit-level heterogeneity and poor control over potential factors that may lead to confounding. Moreover, there are diminishing returns from increasing sample size: in general, the standard error in an estimate tends to decrease with $\frac{1}{\sqrt{n}}$, where $n$ is the number of observations.  This means that when the sample size is already quite large, each additional observation has a vanishingly small benefit to the precision of estimation. `Pilot design' approaches to observational studies consider how researchers can thoughtfully use plentiful data to plan a stronger study design. Said another way: how can investigators use the strengths of abundant observational data to anticipate and mitigate against its weaknesses?

Unlike in a randomized experiment, any claim of a causal effect based on observational data must address the possibility of bias stemming from non-random treatment assignment. 
Matching methods attempt to adjust for this bias by recreating a randomized experiment, grouping treatment and control subjects in a way that balances the distributions of observed covariates \citep{stuart2010matching}. 
However, matching does not guarantee balance in the unobserved covariates; practitioners typically carry out their analyses making the unverifiable assumption that all the relevant covariates have been observed.
Subsequently, sensitivity analyses can quantify how robust the results -- derived under assumptions of strong ignorability -- are to the presence of unobserved bias (see section \ref{notation} and work by Rosenbaum \citep{rosenbaum2005sensitivity, rosenbaum2010designbook}).
 
Propensity score matching and Mahalanobis distance matching, the two most popular matching metrics for causal inference, can be viewed as recreating two different kinds of experimental designs: the completely randomized experiment and the highly blocked randomized experiment, respectively \citep{king2016propensity}. 
A completely randomized experiment will still lead to an unbiased estimate of the treatment effect. However, a highly blocked randomized experiment -- which attempts to control for sources of nuisance variation by assigning subjects to more homogeneous blocks -- will tend to be more statistically efficient \citep{box2005statistics}.   This is the stance taken by King and Nielson in their provocative paper ``Why Propensity Scores Should Not be Used for Matching.'' \cite{king2016propensity}
In randomized experiments, increasing the sample size and reducing the within-block heterogeneity play interchangeable roles, since both reduce the sampling variability of the treatment effect estimate. Surprisingly, in observational studies, reducing heterogeneity within matched sets has an \textit{added} benefit over increasing sample size: reduced sensitivity to unobserved bias. In less heterogeneous observational studies, stronger unobserved confounding would need to be present to explain away the same observed effect \citep{rosenbaum2005heterogeneity}. 

In light of this result, it would seem that Mahalanobis distance matching, which attempts to match on all the observed covariates at once, might be preferable to propensity score matching, which only attempts to match on a score involving covariates that are predictive of receiving the treatment.
However, Mahalanobis distance matching is known to suffer from the curse of dimensionality. Since in large dimensions all observations are in some sense ``far away'' from each other, this method is forced to compromise on poorer quality matches \citep{abadie2006large}.

The less commonly used prognostic score, formalized by Hansen \citep{hansen2008prognostic}, models the subject's expected response had they not received treatment, based on the observed covariates. Matching on this quantity reflects the experimental ideal of controlling for systematic variation in the response under control settings, which should reduce within-matched set heterogeneity compared to propensity score matching.
Analogous to the propensity score, the prognostic score can reduce the covariates to a scalar quantity and -- under suitable assumptions -- results in a form of covariate balance which leads to valid inference.  The primary difference is that the prognostic score weights covariates based on how predictive they are of the outcome, while the propensity score emphasizes covariates based on how predictive they are of the treatment assignment.

Unfortunately, matching on the prognostic score poses new challenges. First, when using the propensity score, it is straightforward to assess whether covariate balance holds for the observed data. In contrast, we cannot assess prognostic score balance with certainty since we do not observe the counterfactual responses of the treatment group had they not received the treatment. To help mitigate this issue, Hansen recommends matching jointly on the propensity and prognostic score. 
Leacy and Stuart \citep{leacy2014joint} begin to explore this approach in simulation studies, suggesting that treatment effect estimation is improved from matching jointly on both scores.  Importantly, they find that methods which use both the propensity and the prognostic score maintain good performance even when one of score models is incorrectly specified. This observation is reaffirmed by Antonelli et al. \citep{antonelli2018doubly}, who demonstrate that, under certain assumptions, matching jointly on propensity and prognostic scores fit using a lasso model is doubly robust -- i.e. estimation from this approach is consistent as long as at least one of the score models is correctly specified. 

However, each of these prior studies of joint matching on propensity and prognosis focus on approaches which fit the prognostic model on the entire control reserve \cite{antonelli2018doubly, leacy2014joint}. An important outstanding challenge in the application of prognostic scores is that same-sample estimation of the prognostic model tends to cause overfitting to the control group, which may bias estimation of treatment effects \citep{hansen2008prognostic, hansen2006report, antonelli2018doubly}. In the experimental setting, Abadie et al. voice closely-related concerns: they find that stratification using a prognostic model fit on the full sample of experimental controls yields biased estimates of treatment effect \cite{abadie2018endogenous}. Indeed, the central double-robust estimation result from Antonelli et al. \textit{requires} that separate samples are used for fitting the score models versus estimating the treatment effect, \cite{antonelli2018doubly} yet the details and trade-offs of such an approach have not yet been directly considered.

To this end, we propose the use of a ``pilot design,'' in which some observations are expended in a principled way to improve the design of a study. Importantly, while this paper centers around an example prognostic score approach, this is not the only application of a pilot design.  Other authors from various fields of the literature have proposed the use of some form of sample splitting to inform a variety of design considerations, including the selection of the primary outcome \cite{heller2009split}, the characterization of an instrumental variable \cite{angrist1992effect, inoue2010two, zhao2019two}, and even the development of the entire analysis plan \cite{fafchamps2017using}. The notion of ``honesty'' from Wager and Athey engages a related data splitting technique for analyses of heterogeneous treatment effects \cite{wager2018estimation} The primary appeal of pilot design approaches is that they enable the use of some ``post-exposure'' information on a subset of the data in a way that maintains the careful separation between the study design and the study analysis.

Section \ref{notation} establishes notation and provides an overview of a common sensitivity analysis considered in this text. Section \ref{methods} describes the pilot design in general and suggests a pilot design for joint matching on the prognostic and propensity scores, then outlines some motivating mathematical results.  This section additionally introduces the ``Assignment-Control (AC) plot,'' an application of prognostic score in the form of a useful visualization tool for observational studies. Sections \ref{Simulations} and \ref{results} summarize simulations comparing the prognostic score pilot design approach to Mahalanobis distance and propensity score matching. Section \ref{considerations} considers the benefits and trade-offs of using this and related pilot designs, and section \ref{discussion} offers some concluding remarks.

\section{Notation and Background}\label{notation}
We adopt the Neyman-Rubin potential outcomes framework, in which a sample is described by
\[\dataset=\{(X_i, T_i, Y_i)\}_{i=1}^n,
\]
where individual $i$ has covariates $X_i$, treatment assignment $T_i$, and potential outcomes function $Y_i$. We will suppress the $i$ index when context permits. We take $T_i = 1$ to represent that individual $i$ was assigned to the treatment group and $T_i = 0$ to represent the control assignment. We define $Y_i(T_i)$ to be the outcome of the individual under treatment assignment $T_i$. 

The fundamental challenge of causal inference is that we cannot observe both $Y_i(0)$ and $Y_i(1)$.  Instead, we observe only one outcome, $Y_i = T_i Y_i(1)+(1-T_i)Y_i(0)$ for a particular observation $i$.  Thus, rather than estimating the individual level effect, researchers may target $\text{E}[Y(1) - Y(0)|i \in sample]$, the sample average treatment effect (SATE\cite{robins1988confidence}).  As in a randomized controlled trial, the SATE can be related to the population average treatment effect (PATE\cite{balzer2016targeted}) through either a known sampling frame that gave rise to the sampled data set, or through careful considerations of the characteristics of the transportability of the treatment.  In practice, it is also common to estimate the sample average treatment effect among the treated (SATT\cite{sekhon2020inference}), which focuses on estimating what would happen if the observational units were moved to the control level. We take the SATT as our target estimand in this paper; other versions of pilot matching could be used to target the PATT (population average treatment effect among the treated) or PATE and would incorporate other analyst-directed information (e.g., key characteristics of transportability\cite{tipton2014generalizable} such as the structure of the heterogeneity of the treatment effect).

The propensity score is defined as $e(X)=P(T=1|X)$.  Interest in the propensity score primarily stems from its use as a balancing score\citep{rosenbaum1983central}, i.e.
\begin{equation}\label{balance}
T \independent X \mid e(X).
\end{equation}
That is, for level sets of propensity score, treatment assignment is independent of the measured covariates. Under the assumption of no unobserved confounding, and assuming that there is no possible value of the covariates $X$ for which the probability of treatment is 0 or 1, subclassification on the propensity score allows for the estimation of the treatment effect.

The prognostic score is defined by Hansen \citep{hansen2008prognostic} as any quantity $\Psi(X)$ such that
\begin{equation}\label{pbalance}
Y(0)\independent X \mid \Psi(X).
\end{equation}  In particular, when $Y(0)|X$ follows a generalized linear model, $\Psi(X) = E\left[Y(0) | X \right]$.  The prognostic score is, by definition, a balancing score. In other words, for level sets of the prognostic score, the potential outcome under the control assignment is independent of the observed covariates.  Under the assumption that there is no unobserved confounding and that there is overlap between the treated and control groups for all values of $\Psi(X)$, prognostic score balance also allows for estimation of the treatment effect.  The assumptions required to estimate the treatment effect by subclassification on the propensity or prognostic score are slightly different, as discussed in detail by Hansen \citep{hansen2008prognostic}.

\subsection{Sensitivity to Unobserved Confounding} \label{gamma_sensitivity}
Rosenbaum \citep{ rosenbaum2005sensitivity, rosenbaum2010designbook} discusses a common sensitivity analysis to examine a study's vulnerability to unobserved confounding.  In this approach, the researcher supposes that assignments to treatment or control within an observational sample are actually determined in part some unmeasured confounding factor, and they attempt to estimate what strength of unmeasured confounding would have to be present for the results of the study to come into question. In specific, they hypothesize that there is some unobserved confounder at play which causes certain individuals' odds of treatment to be $\Gamma$-times greater than other individuals with the same observed baseline covariates.  If $\Gamma = 1$, individuals with the same observed baseline covariates all have the same odds of treatment, as they would in a  randomized controlled experiment.  Greater levels of $\Gamma \geq 1$ mean that, for individuals with the same observed baseline covariates, there is a greater departure in unobserved covariates which govern treatment assignment. One outcome of a sensitivity analysis is an estimate of the greatest possible value of $\Gamma$ which could be at play before the conclusions of the study could change.  Studies whose results would be reversed for $\Gamma \approx 1$ are considered extremely sensitive to imbalances in unobserved baseline covariates, whereas studies whose results would only be reversed in the presence of very large values of $\Gamma$ would be considered more robust.

One interesting motivation for reducing the prognostic heterogeneity between matched individuals in observational studies lies in the additional power provided in sensitivity analyses for unobserved confounding.  Rosenbaum\citep{rosenbaum2005sensitivity} discusses the relative benefits from increasing the sample size of an observational study versus decreasing the heterogeneity of the sample. While both efforts will diminish the variance of the estimator, decreasing the unit heterogeneity in an \textit{observational study} has the added effect of making the conclusions of the study more robust to having their conclusions explained away by unobserved confounding (i.e. $\Gamma$). Intuitively, we imagine in our sensitivity analyses that we have some hidden adversary who, with a strength of $\Gamma$, shifts the treatment probabilities of certain individuals in order to bias our results. If matched individuals are wildly different in terms of their likely outcomes in the absence of treatment (i.e. large vertical distance in the AC plot in Figure 1), then our imagined adversary has greater ability to manufacture spurious results, even when $\Gamma$ is small.  In contrast, if we select matched sets of individuals that are more similar in terms of their likely potential outcomes, the results are less easily explained away.

It is important to not confuse what this form of sensitivity analysis can, and cannot, achieve. Without study design features that provide additional information about the treatment and assignment to treatment (e.g., an instrumental variable, a forcing variable for a discontinuity design, a known null outcome), we cannot estimate the impact of bias arising from unobserved covariate imbalance between the contrast groups. Instead, a gamma sensitivity analysis is a statement about the hypothetical magnitude of biasing treatment assignment that would be required to change the study's conclusions.  Rosenbaum \cite{rosenbaum2010designbook}, compares this concept to that of statistical power: As researchers, we would like to have a well powered study design, but this does nothing to guarantee that there is any underlying difference to be found.  Rather, a well-powered study increases the probability that -- in the situation that the null hypothesis \textit{is} false -- we will come to the correct conclusion.  Likewise, study designs which are better powered in their sensitivity analyses will increase the probability that -- when we do correctly ascertain a treatment effect -- we find the results difficult to explain away with unmeasured confounding. 

\section{Methods}\label{methods}
\subsection{Motivation - The Pilot Design}\label{motivation}

Careful causal inference literature makes a distinction between the ``design phase'' and the ``analysis phase'' of a study \citep{rubin2008design}. During the design phase, all outcome data from the study is masked or unknown to the researcher, while important decisions about data collection, pre-processing, inclusion, and exclusion are made. Once the design phase is completed and fixed, the analysis phase begins. At this moment, the outcomes are revealed, and an estimation of causal effect is performed. This separation of design and analysis protects the integrity of the research by insulating the study design against any influence by the eventual study results. In prospective randomized trials this separation is often enforced by time: the outcome does not come into existence until after the study design becomes fixed. In observational studies, the researcher is responsible for enforcing this separation.

In the experimental setting, there is one case in which the researcher may be allowed to observe some post-exposure information during the design phase; this occurs when a pilot study is run. In this case, before running the experiment, researchers set aside a fixed amount of resources to conduct a smaller ``pilot'' study. The outcomes from the pilot study influence the subsequent study design, but the individuals in the pilot study do not reappear in the final experiment, thereby upholding the separation between the study design and analysis. 

Extending this concept to the observational setting, we propose the \textit{pilot design} for the observational study. Using this approach, researchers set aside a sub-sample of their observations shortly after data collection (the ``pilot set'').  Post exposure information in the pilot set can be used by the researcher to explore improvements to the study design, and these observations are subsequently excluded from the data used in the final analysis (the ``analysis set''). In this paper, we chose to focus on the use of a pilot design to obtain prognostic information. That is, we examine the allocation of a pilot set to fit a prognostic model  (see section \ref{algorithm} for a specific allocation scheme). This prognostic model is then used to estimate prognostic scores for the observations in the analysis set, in order to select higher quality (less heterogeneous) matches in the analysis set. 

However, prognostic score estimation is not the only potential use of the pilot design. For example, the pilot design may be used in the characterization of an instrumental variable (analogous, for example, to when a pilot study is used to understand compliance behavior and recruitment for a larger randomized controlled trial), or may be useful for model fitting when using an inverse weighting technique \cite{angrist1992effect, inoue2010two, zhao2019two}. In other cases, a pilot design approach may aid in the selection of the primary outcome \cite{heller2009split} or analysis plan \cite{fafchamps2017using}.  When data are plentiful in an observational study, allocating these data ``resources'' towards improving the study design in such ways may prove more useful than reserving all data for the main analysis.

\subsection{Assignment-Control Plots and Matching}\label{fmplots}

Estimation of the prognostic score enables a useful data visualization. Figure 1 visualizes the objectives of three different matching approaches in an \textit{Assignment-Control (AC) plot}. We imagine a scenario in which each individual in our data set is represented in a reduced space of only two metrics: a measure of the covariates determining the expected treatment assignment ($\phi(X_i)$), and a measure of the covariates determining the expected outcome under control assignment ($\Psi(X_i)$)\footnote{Other dimensions -- such as one for instrumental variables, or one for expected treatment outcome under treatment -- might be included as useful extensions to these plots}. In this way, reducing variation in the horizontal direction equalizes the probability of treatment between compared individuals ("assignment") while reducing variation in the vertical dimension imposes balance on covariates important to the outcome ("control") \citep{rosenbaum2005heterogeneity}. These are two (often inter-related) features which are directly relevant to our matching: $\phi(X_i)$ (propensity score) similarity between compared observations reduces bias, and $\Psi(X_i)$ (prognostic score) similarity reduces bias as well as variance, while increasing the power of sensitivity analyses of unobserved confounding.  The example given below illustrates how such a representation may help a researcher intuitively grasp considerations of a causal question and data set. 

Optimal Mahalanobis distance matching (Figure 1A), pairs individuals who are closest in the full covariate space. However, since not all covariates are important for prognosis or treatment assignment, individuals who are close in the full covariate space may be relatively distant when considered in terms of $\phi(X_i)$ and $\Psi(X_i)$. Propensity score matching (Figure 1B) aims to pair individuals who are close in the covariates important for treatment assignment, $\phi(X_i)$, but not for prognosis, $\Psi(X_i)$. This matching will reduce bias in the estimation of causal effect compared to the unmatched data set, but will lose much of the protection from variance and unobserved confounding conferred when the matched individuals are prognostically similar. In contrast, matching jointly on $\phi(X_i)$ and $\Psi(X_i)$, as proposed by Hansen \cite{hansen2008prognostic}, seeks to pair individuals who are close together in the reduced feature space. This optimizes for both desirable types of covariate balance: prognostic and propensity.

\begin{figure}[h]
\centering
  \includegraphics[scale = 0.8]{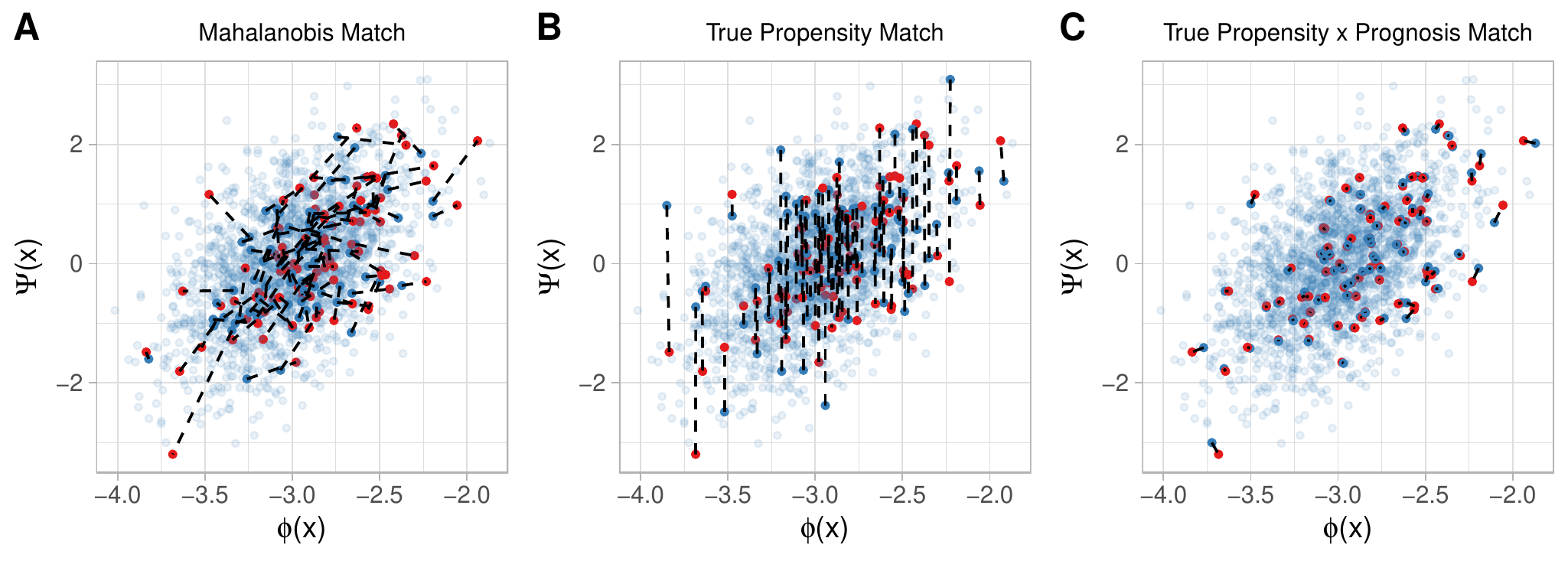}
  \caption{AC plots with matching assignments. The horizontal axis, $\phi(X)$, shows variation important for determining treatment assignment (log-propensity score), while the vertical axis, $\Psi(x)$, shows variation important for determining the outcome in the absence of treatment (prognostic score). Blue dots represent control individuals, red dots represent treated individuals, and dotted lines connect matched pairs. Each plot shows optimal 1:1 matches of a simulated data set based on Mahalanobis distance in the whole covariate space (A), true propensity score (B), and Mahalanobis distance in the feature-reduced space of true propensity score and true prognostic score (C). The data set was generated according to the simulation set up in section \ref{setup} with $\rho = 0.5$.}
 \end{figure}

\subsection{Prognostic Pilot Matching}\label{algorithm}

We propose the following \textit{pilot matching} design for constructing and applying the prognostic score in the study design phase (Algorithm 1). The main steps are:
\begin{enumerate}
  \item Fit a propensity model, $\hat{\phi}$, on the entire data set, $\mathcal{D}$.
  \item Separate the data into a held-aside pilot sample $\mathcal{P}$ and an analysis sample, $\mathcal{D}'$
  \item Fit a prognostic score model, $\hat{\Psi}$, using outcome information from only $\mathcal{P}$.
  \item Perform Mahalanobis distance matching on $\mathcal{D}'$ based on the prognostic and propensity scores estimated from $\hat{\phi}$ and $\hat{\Psi}$.
\end{enumerate}

One method for doing so is suggested in Algorithm 1. In this approach, the pilot set is selected by first constructing a 1:2 Mahalanobis distance matching of each treated individual (denoted $A_1,...,A_{n_T}$, where $n_T$ denotes the number of treated individuals) to a set of two control individuals (denoted $B_1,..., B_{n_T}$). Then from each pair of matched controls ($B_i$), we select one individual at random ($b_i$) to place into our pilot set. This gives a pilot set, $\mathcal{P}$ of size $n_T$, which is used to estimate the prognostic model and then removed from the analysis. The preliminary 1:2 Mahalanobis distance matching ensures (1) that the individuals in the prognostic set are relatively close in the covariate space to the treated individuals, minimizing extrapolation of $\hat{\Psi}$, and (2) that not all individuals close in the covariate space to the treated individuals are removed from $\mathcal{D}'$, because this might sacrifice too many potentially high-quality matches.

\begin{algorithm}
\caption{Prognostic Pilot Matching}\label{alg:prog}
\begin{algorithmic}
 \State \textbf{Input:}$\mathcal{D}=\{(X_i,T_i,Y_i)\}_{i=1}^n$, fixed treated to control ratio $k$.
 \State Fit a linear propensity model $\hat\phi$ using $\mathcal{D}$.
 \State Construct a $1:2$ Mahalanobis distance matching $(A_1,\hdots, A_{n_T};B_1,\hdots, B_{n_T})$.
 \For{$i \gets 1$ \textbf{to} $n_T$}
    \State $b_i\gets$ a control chosen uniformly at random from $B_i$. 
 \EndFor
 \State $\mathcal{P} \gets \{b_i\}_{i=1}^{n_T}$
 \State $\mathcal{D}'\gets \mathcal{D} \cap \mathcal{P}^C$
 \State Fit a linear prognostic model $\hat \Psi$ using held out controls $\mathcal{P}$.
  
 \State \textbf{Output:}$1:k$ Mahalanobis distance matching of $\mathcal{D}'$ using $\hat\phi$ and $\hat \Psi$.
\end{algorithmic}
\end{algorithm}

There are several alternative approaches to carry out steps 1-4; we do not intend to prescribe a single method for doing so. For example, at steps (1) and (3), a researcher may select from any number of models to fit. In cases where the covariate space is especially large, sparsifying models such as the lasso may be appropriate (Supplementary Figure 1 and Antonelli et al.\citep{antonelli2018doubly}). In step (4), the researcher may choose between pairmatching, 1:k matching (for example, in Figure 1), full matching (Figure 5), or some other sub-classification method.

There are a variety of options for the allocation of the pilot set in step (2). For example, systematically selecting the lower quality match from the set of matched controls might ensure that high quality matches to the treated individuals are left behind for the analysis set. When data are especially plentiful or the prognostic model is especially elusive, one might perform a preliminary matching of 1:4 or more (rather than 1:2) and select more than a single control individual from each matched set to allocate more data to the pilot set. If the sample is so large that a preliminary Mahalanobis distance matching on the entire data set would be too computationally taxing, the pilot set might be selected at random, either uniformly or while stratifying by important covariates (e.g. sex, smoking status, age group). Each of these decisions should ultimately be made based on the specific data set and research question at hand.

\subsection{Mathematical Results}\label{maths}

In this section we include motivating mathematical results supporting the joint usage of propensity and prognostic score information in matching. First, we give forms of the bias, variance, and MSE of a straightforward pair-matching effect estimator in terms of the prognostic score, as an illustration of how the moments of the estimator can be bounded by the prognostic score differences (for any matching scheme). Second, we review and slightly extend the results of Antonelli et al. \citep{antonelli2018doubly}, to give the asymptotic consistency of a prognostic pilot matching estimator. In particular, asymptotic consistency is achieved as long as at least one of two models (for propensity and prognostic score) is correctly specified. This implies that prognostic pilot matching designs are doubly robust.

\subsubsection{Properties of a simple matching estimator in terms of prognosis}

Suppose we observe $n_T$ treated individuals (indexed by $i$), and suppose each treated individual is matched to exactly one control individual (using any matching scheme). Let $j_{(i)}$ represent the index of the control individual matched to the $i$-th treated individual.  Finally, suppose we additionally assume the following:
\begin{enumerate}
    \item The outcome is continuous, and for all individuals, $Y(0) = \Psi(X) + \epsilon,$ where $\epsilon \iid N(0, \sigma^2)$
    \item $\epsilon$ is independent of $\Psi(X)$
    \item The differences in outcomes between matched pairs, $D_i := Y_i - Y_{j_{(i)}}$, are independent.
\end{enumerate}

\begin{theorem}\label{thm1}
Assume the statements above and suppose $\tau$ is a constant additive treatment effect. Let $\Psi(X_i) - \Psi(X_{j_{(i)}})$ denote the difference in prognostic score for a randomly selected treated individual and its matched control.  If we estimate the treatment effect via $\hat{\tau} := \frac{1}{n_T}\sum_{i = 1}^{n_T} D_i$, then 
\begin{equation}
    Var(\hat{\tau}) = \frac{4\sigma^2}{n_T} +\frac{1}{n_T} Var\left(\Psi(X_i) - \Psi(X_{j_{(i)}})\right),
\end{equation}

\begin{equation}
    Bias(\hat{\tau}) =  E\left[\Psi(X_i) - \Psi(X_{j_{(i)}})\right],
\end{equation}

and
\begin{align}
    MSE(\hat{\tau}) &= \frac{4\sigma^2}{n_T} +\frac{1}{n_T} Var\left(\Psi(X_i) - \Psi(X_{j_{(i)}})\right) +\left(E\left[\Psi(X_i) - \Psi(X_{j_{(i)}})\right]\right)^2 \\
    &\leq \frac{4\sigma^2}{n_T} + E\left[\left(\Psi(X_i) - \Psi(X_{j_{(i)}})\right)^2\right]
\end{align}
Moreover, if assumptions 2 and 3 do not hold, the right hand side of (3) and (5) become upper bounds, and (4) and (6) still hold.

\end{theorem}

Theorem 1 expresses the bias, variance, and MSE of a simple matching estimator in terms of the prognostic score. A derivation is supplied in the Appendix. Note that this theorem holds regardless of the method used to select the matched pairs.  In particular, we find that the variance in the matching estimator is bounded by the \textit{variance} in the difference between the prognostic scores. This means that the variance in estimation can be small even when matched individuals are very prognostically different, as long as this difference in prognostic score is relatively \textit{consistent}  across all matched pairs. However, this situation is still not desirable because a large mean difference in prognostic score within pairs will increase bias, as indicated by Equation (4).  

Equation (5) combines these results, reiterating how minimizing both the mean difference and the variance in the difference in prognostic scores across matched pairs gives a reduction in MSE. Finally, (6) shows how the MSE of the estimator can be bounded by the mean squared variation in prognostic score between matched pairs (although this bound is fairly loose). An interesting insight from this theorem is that prognostic score based matching methods decrease variance not just by reducing the prognostic score distances between matched individuals, but by making those distances more \textit{consistent}.

\subsubsection{Doubly Robust Pilot Matching}

Theorem 1 from Antonelli et al.\citep{antonelli2018doubly} shows that the average treatment effect is identifiable when conditioning jointly on the propensity and prognostic score. In specific, this identifiability holds as long as at least one of the two score models is correctly specified. This helps to explain the simulation results of Leacy and Stuart \citep{leacy2014joint}, in which matching methods which used the propensity and prognostic score jointly performed well even when one model was incorrectly specified.

In 2005, Bang and Robins\citep{bang2005doubly} introduced a notion of ``doubly robust'' estimation. This term has come to refer to any estimator which combines an outcome model with a model for the probability of treatment in such a way that the estimator is consistent or (more informally) ``correct'' as long as at least one of the models is correctly specified. Antonelli et al.\citep{antonelli2018doubly} propose a doubly robust matching estimator (DRME), which uses a lasso regression to estimate $\hat{\phi}$ and $\hat{\psi}$. Theorem 2 of Antonelli et al. shows that the lasso estimator is indeed consistent under suitable regularity conditions, provided at least one of the models is correctly specified. Below, we state a slightly generalized version of that theorem which allows for a more broad class of doubly robust matching designs (i.e. designs in which $\hat{\phi}$ and $\hat{\psi}$ can be estimated by methods other than the lasso). In this form of the theorem, we clarify how the rate of convergence for the estimator from the pilot design depends on: (1) the sample size of the pilot data set (2) the sample size of the analysis data set, and (3) the specific method selected to fit $\hat{\phi}$ and $\hat{\psi}$.

Let $n_{\mathcal{D}'}$ and $n_\mathcal{P}$ be the sample sizes of the analysis and the pilot data sets, respectively. Following the notation of Antonelli et al.\citep{antonelli2018doubly}, let $\hat{\theta}$ represent a vector of the estimated parameters for the propensity and prognostic models ($\hat{\phi}$ and $\hat{\psi}$), and, for a given $\hat{\theta}$, let $Z = \left(\hat{\phi}(X), \hat{\Psi}(X)\right)$. Additionally, let $\tilde{\theta}$ represent the probabilistic limit of $\theta$ (Note that this may not be the true model parameters, since the models for $\phi$ and $\psi$ may be mis-specified). In keeping with Abadie and Imbens\citep{abadie2006large} and Antonelli et al.\citep{antonelli2018doubly}, suppose the $M$ matches for the $i^{th}$ individual with score model parameters $\theta$ are given by

\begin{equation}
   \mathcal{J}_M(i, \theta) = \left\{j = 1, ... , n_{\mathcal{D}'}: T_j = 1- T_i, M \geq \sum_{l:T_l = 1-T_i}I\left(\lVert Z_i - Z_l\lVert < \lVert Z_i - Z_j\lVert\right)\right\},
\end{equation}

\noindent where $I$ is an indicator variable and $\lVert\cdot\lVert$ is Mahalanobis distance\footnote{This is slightly different than the matching scheme used in this simulation study and in many study designs, in that each individual (both treated and control) is matched with the $M$ nearest individuals of the opposite treatment assignment in the analysis set, and individuals may be used more than once. In practice, many studies use matching without replacement, which we consider here in simulation.}.

Finally, let $\tau(\theta)$ represent the estimator obtained from the matching given by $\mathcal{J}_M(i, \theta)$:

\begin{equation} \label{eqtn-tau definition}
    \tau(\theta) = \frac{1}{N_{\mathcal{D}'}}\sum_{i = 1}^{N_{\mathcal{D}'}}(2T_i - 1)\left(Y_i - \frac{1}{M}\sum_{j \in \mathcal{J}_M(i, \theta)}Y_j\right).
\end{equation}

\begin{theorem} \label{thm2}(Generalization of Antonelli et al. \citep{antonelli2018doubly})
Let $\tau(\theta)$ be the with-replacement "radius" matching estimator described in \ref{eqtn-tau definition}, and suppose that $\hat{\theta}$ is estimated on the pilot data set by some method such that, $\lVert\hat{\theta} - \tilde{\theta}\lVert$ converges in probability to 0 at a rate $R(n_\mathcal{P})$. Under regularity conditions described in the supplement (including no effect modification), and assuming that at least one of the models for propensity or prognosis is correctly specified, then,
\begin{equation}
    \tau - \tau(\hat{\theta}) = O_p\left(n_{\mathcal{D}'}^{-\frac{1}{2}}\right) +  o_p\left(R\left(n_\mathcal{P}\right)\right)
\end{equation}

If both $n_\mathcal{P}$ and $n_{\mathcal{D}'}$ are allowed to go to infinity as the sample size increases, and if $R(n_\mathcal{P})$ is bounded as $n_\mathcal{P} \rightarrow \infty$, then we have that $\tau(\hat{\theta})$ is a consistent estimator for $\tau$.
\end{theorem}

This result is an almost immediate generalization from the proof for theorem 2 in Antonelli et al.\citep{antonelli2018doubly}; however a more specific explanation can be found in the supplement. 

An explanation of the probabilistic Big O notation used in this theorem can be found in Van Der Vaart, \citep{van2000asymptotic} among other places. Intuitively, the first term on the right hand side of (10) describes how the error in estimation depends on the sample size of the analysis set: As the size of the analysis set increases, the number and quality of matches will increase, allowing for better estimation. The second term on the right hand side describes how the error depends on the sample size of the pilot set: As the sample size of the pilot set increases, we will fit more accurate models for propensity and prognosis, allowing us to identify the higher quality matches from the analysis set.

This form of the theorem provides for doubly robust matching estimation from a prognostic pilot matching design, provided any suitable choice of method for estimating $\theta$. Additionally, it is important to note that the sample sizes of both the pilot data set and the analysis data set must go to infinity for consistency, yet the choice of the specific method used to partition the pilot and analysis data sets can be left (somewhat) to the researcher - provided the selection of the pilot set is appropriate for consistent estimation of $\tilde{\theta}$.  A deeper discussion of the practical design choices for a pilot matching procedure can be found in \ref{algorithm}. Note that a substantial implicit assumption of any doubly robust theorem is that there \textit{is} some true underlying score model (propensity or prognosis) which holds for the entire collected data set: treated, and control.

\section{Simulations}\label{Simulations}

\subsection{Objective} \label{objective}

There are several design considerations involved in an approach which intentionally removes observations from the analysis set phase. Here, we highlight four, often interacting, considerations when selecting a prognostic pilot matching design or similar approach:

\begin{enumerate}
    \item \textbf{Correlation of treatment and prognosis (\ref{consideration-rho})} - if the treatment and the outcome are tightly correlated, matching on the propensity score may select pairs that are prognostically similar as well.
    \item \textbf{Trade offs in sample size (\ref{consideration-n})} - allocating data for use in the design phase decreases the amount of data available in the analysis phase.
    \item \textbf{Trade offs in match quality (\ref{consideration-quality})} - since not all control observations are equivalently useful (some may be very distant from the treated individuals in terms of important covariates), care must be taken selecting which controls to set aside.
    \item \textbf{Fitting the propensity and prognostic models (\ref{consideration-fitting})} - if the data generating process for treatment assignment and outcome is particularly hard to model, this may influence the usefulness of methods which rely on fitted scores.
\end{enumerate}

We performed a simulation study to assess the performance of the prognostic pilot matching approach detailed in section \ref{algorithm} compared to Mahalanobis distance and propensity score matching.  The following subsection describes the data-generating process used in our simulations, and section \ref{results} gives the results, focusing in section \ref{considerations} on the considerations listed above.

\subsection{Setup} \label{setup}
We compare the performance of propensity score matching, Mahalanobis distance matching, and a prognostic pilot matching approach (described in section \ref{algorithm}) on simulated data.  In each batch of simulations, the fixed control to treatment ratio ($k$) in the matching procedure and the correlation between the true propensity and prognostic score ($\rho$, below) varied across a range of values. Suppose we measure $p$ covariates, so that $X$ is a $p$-dimensional vector.  Let $I_p$ denote the $p\times p$ identity matrix.  The generative model for our main set of simulations is the following:
\begin{align*}
    X_i &\iid \text{Normal}(0,I_p),\\
    T_i &\iid \text{Bernoulli}\left(\frac{1}{1+\exp(-\phi(X_i))}\right),\\
    Y_i &=\tau T_i + \Psi(X_i) + \epsilon_i,\\
    \epsilon_i &\iid N(0,\sigma^2),
\end{align*}
where the true propensity and prognositic scores are given by the linear combinations
\begin{align*}
    \phi(X_i) &= X_{i1}/3-c,\\
    \Psi(X_i) &=\rho X_{i1} + \sqrt{(1-\rho^2)}X_{i2}.
\end{align*}
Importantly, these values ensure that $\rho$ is precisely the correlation between $\phi$ and $\Psi$ (i.e. $\text{Cor}(\phi(X_i), \Psi(X_i)) = \rho$). When $\rho = 0$, the treatment effect is completely unconfounded, since treatment assignments are entirely determined (up to randomness) by variation in $X_{i1}$, and outcome (under the control assignment) is entirely determined by variation in $X_{i2}$. When $\rho = 1$, the problem is highly confounded, since outcome and treatment assignment are both determined solely by variation in $X_{i1}$.

We fix the treatment effect to be constant with $\tau=1$, the number of covariates $p = 10$, and the noise in the outcomes $\sigma = 1$. Each simulation consisted of a data set of total sample size $n=2000$, and simulations were repeated $N=1000$ times. The constant, $c$, in the propensity score formula was chosen such that there were approximately 100 treated observations in each data set (e.g. for a sample size of 2000, the value $c = 3$ was used). We consider $1:k$ matchings of the analysis set for $k=1,\hdots, 10$, and $\rho = 0, 0.1,\hdots, 0.9, 1.0$.  In simulated samples, the Cohen's D standardized difference between the outcomes of the treated and control groups was approximately 0.6 to 1.0 (depending on the value of $\rho$).

We ran the following additional simulations to interrogate the behavior of each method in slightly different scenarios.  The results from these simulations are summarized in Figures 4-6 and Supplementary Figures 1-9.
\begin{itemize}
    \item \textit{Unmeasured confounder} - An additional unmeasured covariate, $U_i \iid \text{Normal}(0,1)$ was added, and the formulas for the true propensity and prognostic scores were given by $\phi(X_i) = X_{i1}/3 + 0.2U_i - c$ and $\Psi(X_i) = \rho X_{i1} + \sqrt{(1-\rho^2)}X_{i2} + 0.2U_i$
    \item \textit{Full matching rather than fixed-ratio matching} - Instead of performing 1:k matching in the final step, full matching was used. As in the $1:k$ matching, the pilot design full match was based on Mahalanobis distance in the combined space of $\hat{\phi}$ and $\hat{\Psi}$. In this case, $N=2000$ simulations were performed to get accurate assessments of performance.
    \item \textit{More uninformative covariates} - The number of covariates was increased from $p = 10$ to $p = 50$. (Note that $X_{i1}$ and $X_{i2}$ remain the only important covariates for treatment assignment or outcome.)
    \item \textit{Smaller sample size} - The sample size was decreased from $n = 2000$ to $n = 1600$. (In this case, $c$ was changed to $2.75$ to keep the number of treated individuals consistent.)
    \item \textit{Worse covariate overlap} - The formula for $\phi$ was changed to $\phi(X_i) = X_{i1} - 10/3$.
    \item \textit{More unpredictable outcomes} - The noise in the outcome was increased from $\sigma = 1$ to $\sigma = 2$.
    \item \textit{Heterogeneous treatment effect} - rather than using a fixed constant treatment effect, the treatment effect was allowed to vary between individuals, with individual effect $\tau(X_i) = 1 + X_{i1}/4$.
    
\end{itemize}
\noindent In addition, we considered the performance of each method when the score models are correctly specified, and when they are fit with a lasso (see \ref{consideration-fitting}). The \verb|glmnet| package was used for lasso fitted models\cite{friedman2010glmnet}.

When using 1:k matching, we estimate SATT and perform sensitivity analyses using the permutation $t$-statistic from the package \verb|sensitivtymv| \citep{rosenbaum2015Rpackages}. When using full matching, we estimate SATT using the package \verb|sensitivtyfull| \citep{rosenbaum2007mestimates}. The accuracy of each matching method was assessed based on the empirical bias, standard deviation, and mean squared error (MSE) of the estimates produced in simulation. For sensitivity analyses, for each simulated study we calculated the largest strength of hypothetical unobserved confounding, $\Gamma$, at which the null hypothesis would still be rejected (with an alpha level of 0.05). Results across batches of simulations are reported in terms of median $\Gamma$. All matching assignments were found using \verb|optmatch|, \citep{hansen2006optmatch}, which makes use of the RELAX-IV algorithm \citep{bertsekas1988relaxation}.  Source code and data files for these simulations are publically available on github (https://github.com/raikens1/PilotMatch).

\section{Results} \label{results}
\subsection{Primary Simulations} \label{primaryresults}

We first consider the main simulation set-up described in section \ref{setup}. In this case, there are approximately 19 control individuals observed for every one treated individual. Figure 2 shows the bias and variance of $1:k$ matching for each method.

\begin{figure}[h]
\centering
    \includegraphics[scale = 0.8]{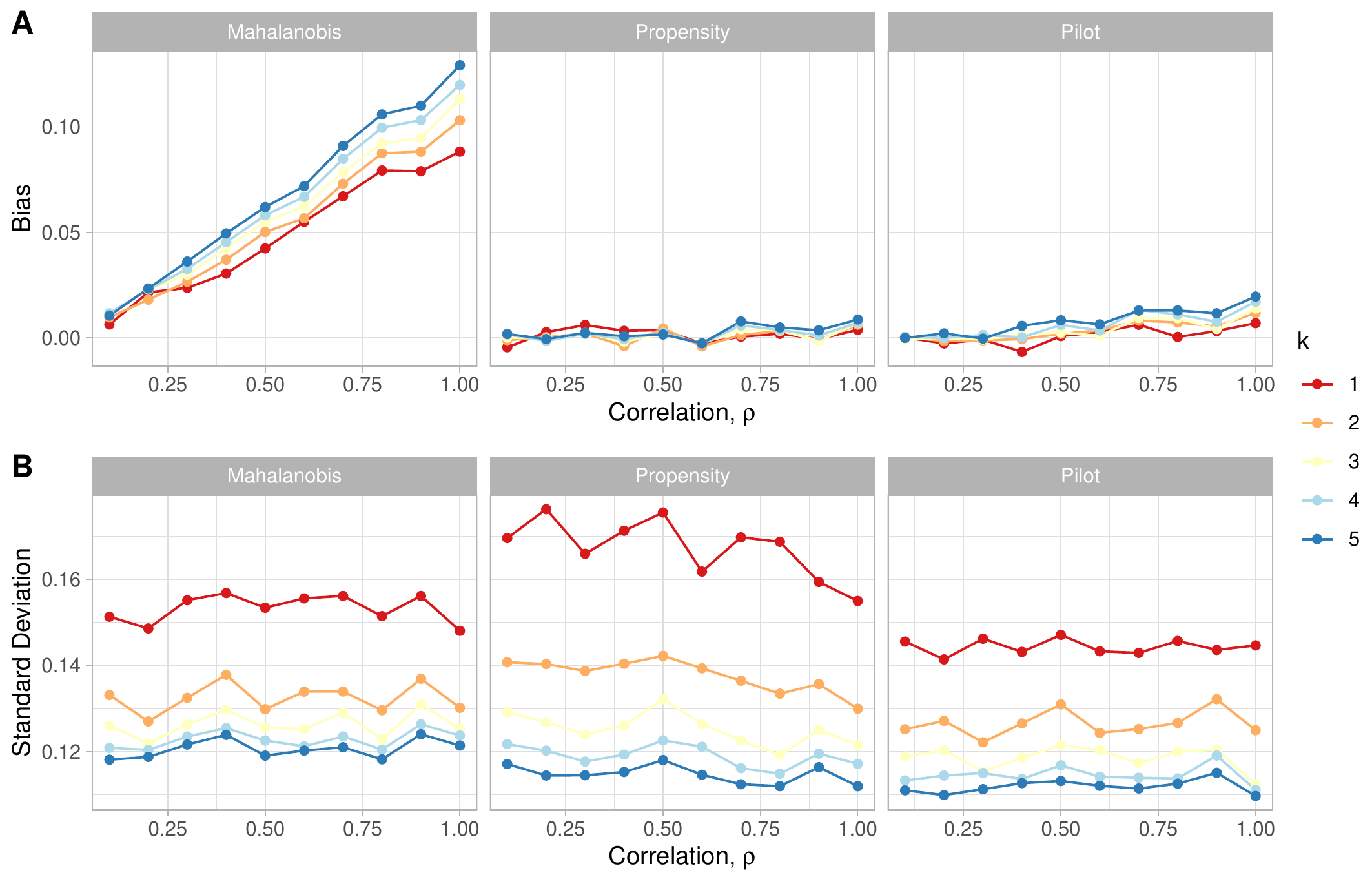}
    \caption{Mahalanobis distance, propensity, and prognostic pilot matching performance in terms of bias (A) and standard deviation (B) for 1:k optimal matching as the correlation between propensity for treatment and prognosis, $\rho$, varies from 0 to 1.}
\end{figure}

First, we observe that bias for all methods increases as the correlation between propensity and prognosis, $\rho$, increases from 0 to 1 (Figure 2A). This is perhaps unsurprising, since $\rho \approx 1$ means that the treatment assignments are highly correlated with the potential outcomes. The Mahalanobis distance matching estimates are by far the most biased for large values of $\rho$, since this method equally weights the informative covariates, ($X_{i1}$ and $X_{i2}$), alongside several other covariates which are unimportant to the problem. This issue is  exacerbated when the number of uninformative covariates is increased (Supplementary Figure 2).  

The bias in propensity and prognostic pilot matching estimates are both much smaller, but there is a minor increase in the bias of prognostic pilot matching when $k$ is large.  This is most likely due to a trade-off in match quality implicit in selecting the pilot set (see section \ref{consideration-quality}).  Importantly, while Figure 2A illustrates trends in bias in a situation with no unobserved confounding, it does not comment on the sensitivity of the design to unobserved confounding (see the discussion of Figure 3, below).

An interesting finding from Figure 2B is that a prognostic pilot matching approach may actually \textit{reduce} the variance in the estimator under the right conditions, in spite of the fact that pilot designs by their nature reduce the size of the sample which may be used for estimation. This underscores the insight that reserving all data for the analysis phase of a study is not always the optimal approach for obtaining reliable point estimates \textit{or} maximizing statistical power. For 1-to-1 matching, the prognostic pilot match approach had 6-20\% lower standard deviation in the estimates compared to propensity score matching, with the greatest difference observed when $\rho$ is close to zero. When $\rho$ is small, the propensity score captures little information about the variation which determines prognosis (that is, it is similar to an instrumental variable). leading to propensity score matches which may be very different in terms of their potential outcomes (Figure 1, see further discussion in section \ref{consideration-rho}).

\begin{figure}[h]
\centering
   \includegraphics[scale = 0.8]{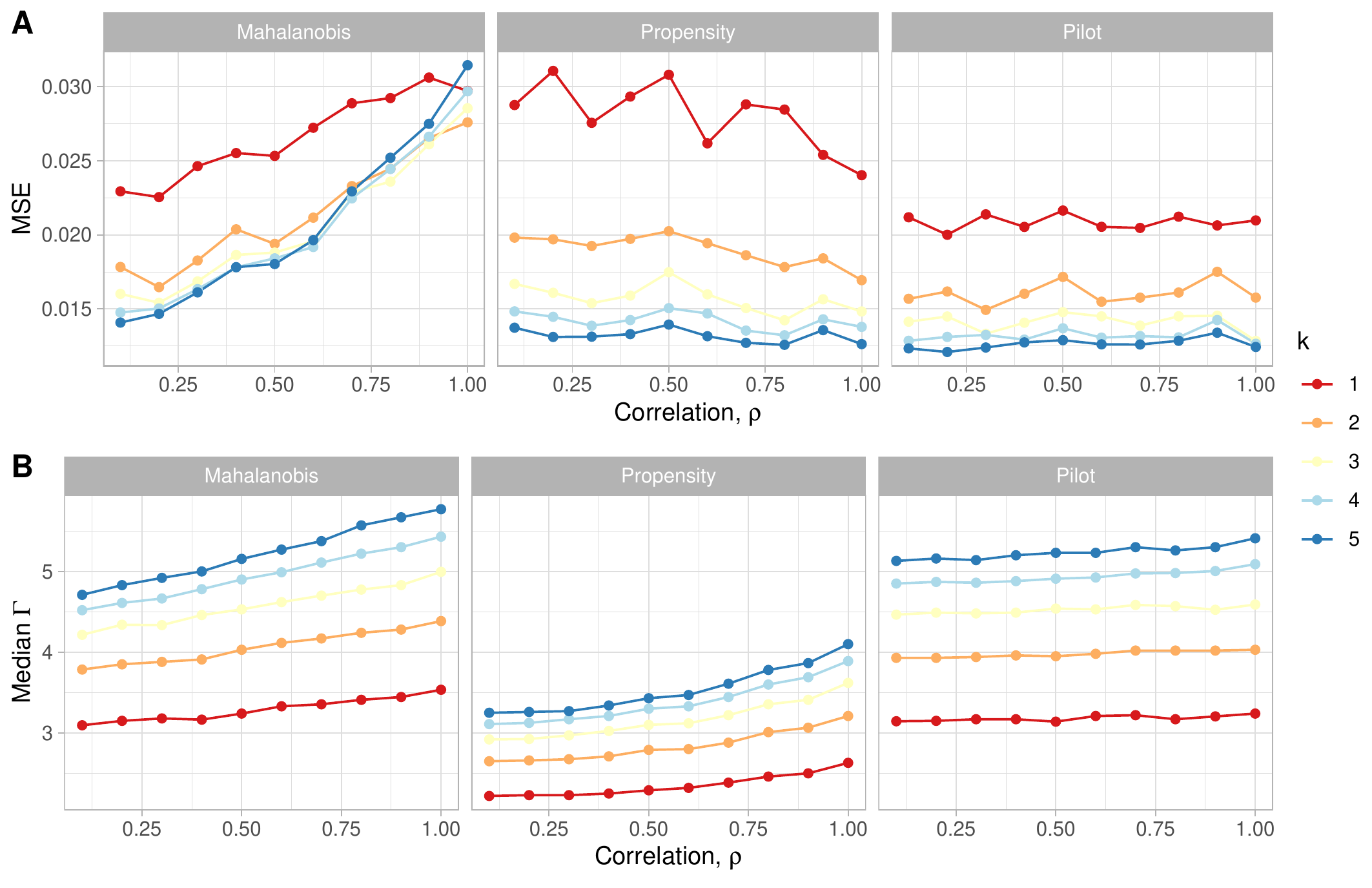}
    \caption{Mahalanobis distance, propensity, and prognostic pilot matching performance in terms of MSE (A) and the median level of confounding, $\Gamma$, at which the study conclusions would no longer be statistically significant (B) for 1:k optimal matching as the correlation between propensity for treatment and prognosis, $\rho$, varies from 0 to 1.}
\end{figure}

In these simulations, prognostic pilot matching generally outperforms both existing methods in terms of mean squared error (Figure 3A). This is because the estimates from this approach are less biased than Mahalanobis distance matching (Figure 2A), and more precise than propensity score matching (Figure 2B).  The percent improvement over propensity score matching is greatest when $k$ is small. For 1-to-1 matching, mean squared error is reduced by 12\% to 36\% compared to propensity score matching and 8\% to 34\% compared to Mahalanobis distance matching (depending on $\rho$).

Figure 3B highlights a unique but often overlooked strength of observational studies in which compared observations have similar potential outcomes: the results are more difficult to explain away with unobserved confounding. For 1-to-1 matching, the magnitude of the effect estimate from propensity score matching could be explained by an unobserved confounder which influenced the odds of treatment by a factor of approximately 2.5.  In contrast, an unobserved confounder would have to increase certain individuals' odds of treatment by 5 fold in order to explain the results when prognostic pilot matching is used. Gamma values from Mahalanobis distance matching seem promising. However, some of this apparent strength is likely attributable to the fact that effect estimates from Mahalanobis distance matching in these simulations tend to be overestimated (Figure 2A). 

It is notable that the large observed difference in power in gamma sensitivity analyses cannot be explained by variance reduction in the estimator alone. In 1:5 matching experiments, the standard deviation in estimation from prognostic pilot matching and propensity score matching is comparable (Figure 2B), yet the gamma values in Figure 3B for these simulations are quite different. By matching on the prognostic score, prognostic pilot matching directly reduces the heterogeneity of matched sets in terms of their likely outcome under the control assignment. This confers a notable benefit: The variance reduction from prognostic pilot matching in Figure 2B could be attained with a sufficient increase in sample size, yet the increased power of sensitivity analyses in Figure 3B is likely a unique contribution of heterogeneity reduction \cite{rosenbaum2005heterogeneity}.

It is important to note that the value of increasing the power of a sensitivity analysis is somewhat nuanced: while strong power in gamma sensitivity analyses decreases the likelihood that perceived results can be explained away by unobserved confounding, this does not mean that unobserved confounding is not influencing the study results. Propensity score methods -- even those augmented by a prognostic score -- still require the absence of unobserved confounders in order to produce unbiased estimates of treatment effect (see section \ref{gamma_sensitivity}).  This apparent contradiction is emphasized by Figure 4, which shows that all three of the matching methods considered here are biased when an unobserved confounder is present, in spite of differing power in gamma sensitivity analyses.

\begin{figure}[h]
\centering
   \includegraphics[scale = 0.8]{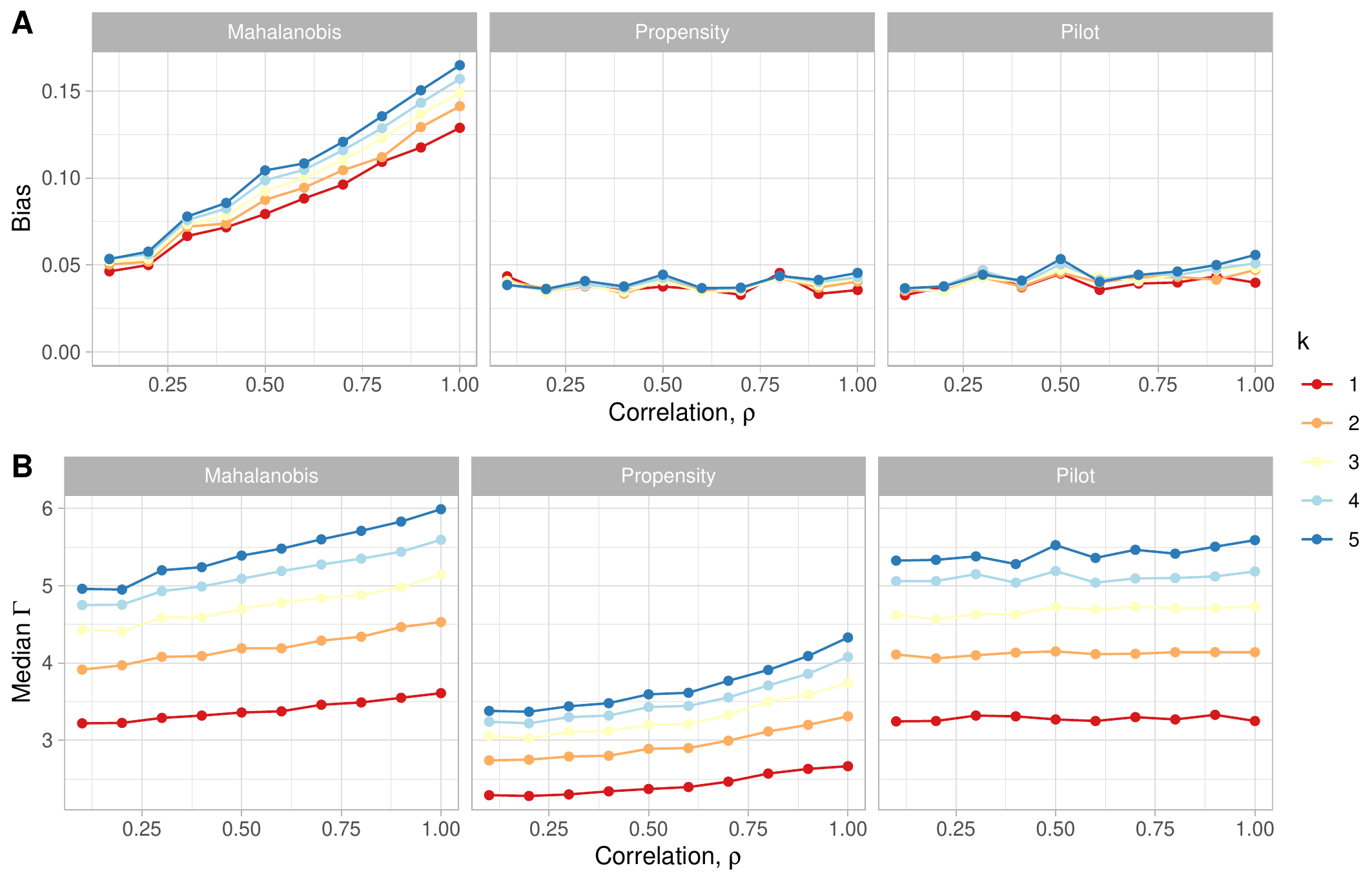} 
    \caption{Mahalanobis distance, propensity, and prognostic pilot matching performance in terms of MSE (A) and the median level of confounding, $\Gamma$, at which the study conclusions would no longer be statistically significant (B) in a scenario with an unobserved confounder.  For simulation details, see \ref{setup}. For variance and MSE results, see Supplementary Figure 3.}
\end{figure}

\subsection{Full Matching} \label{fullmatching}

Prognostic pilot matching designs need not be limited to matching a single treated individual to a fixed number, $k$, of control individuals in the analysis set. Once the prognostic score is built from the pilot data set, the researcher may choose from any number of methods to subclassify or match the analysis set based jointly on the prognostic and propensity scores. To illustrate this point, Figure 5 contains the results of simulations in which the  analysis set was subclassified using full matching \citep{rosenbaum1991fullmatch, hansen2004full}. This method represents a more flexible alternative to fixed 1:k matching, in which the ratio of treated to control individuals is allowed to vary within each matched set. It is similar in spirit to the `variable-ratio' matching applied by Pimentel et al.\citep{pimentel2015variable}. In essence, treated individuals which are similar to many control individuals may be matched with many control individuals, whereas treated individuals which are similar to few control individuals are matched with few control individuals (see Hansen\citep{hansen2004full} or Stuart and Green\citep{stuart2008using} for applied examples). In this way, full matching uses all of the observations in the analysis data set exactly once while often selecting closer matches than fixed $1:k$ matching \citep{hansen2004full}.

\begin{figure}[h]
\centering
    \includegraphics[scale=0.8]{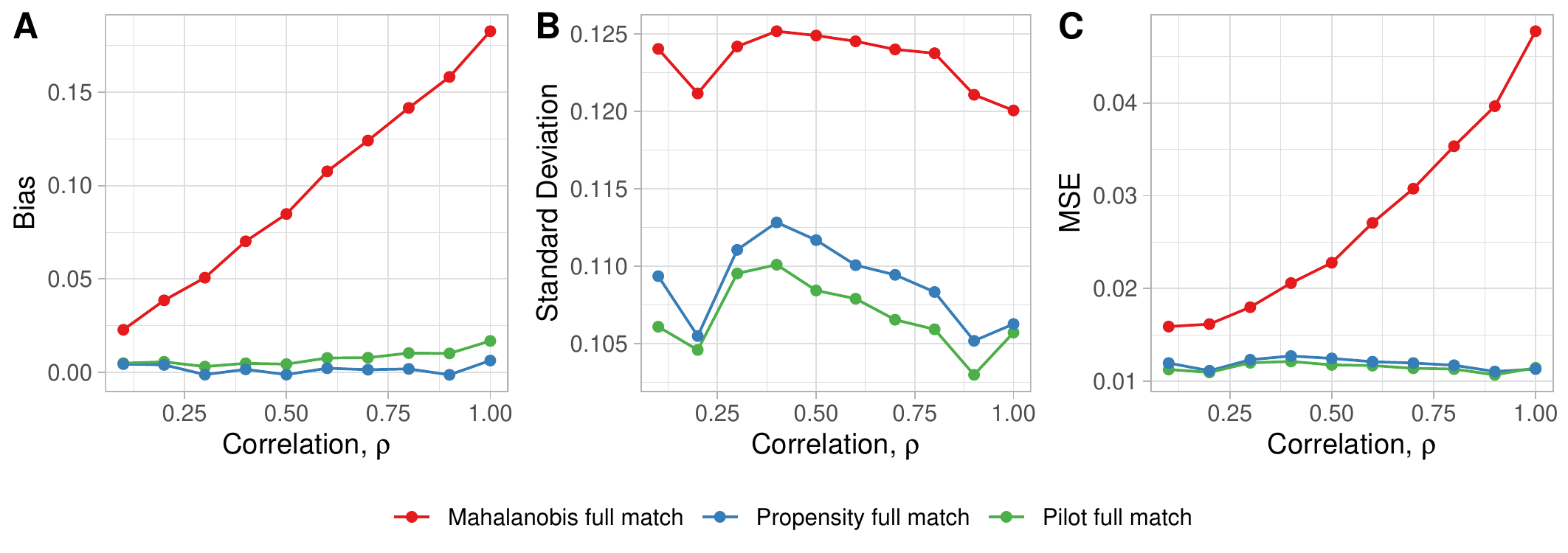}
    \caption{Performance from full matching based on Mahalanobis distance, propensity score, and prognostic pilot matching in terms of bias (A), standard deviation (B), and MSE (C).}
\end{figure}

Figure 5 shows the bias (A), standard deviation (B), and MSE (C) of full matching using Mahalanobis distance matching, propensity score matching, and matching jointly on the propensity and prognosis (`pilot full matching'). We find that full matching using propensity and pilot full matching yield comparable or superior performance compared with $1:k$ matching with a well-selected $k$. Additionally, we find in our simulations that pilot full matching tends to slightly outperform full matching on just the propensity score. These results suggest that full matching jointly on propensity and a prognostic score from a held-aside `pilot' sample may present a promising design strategy for some observational studies.

\subsection{Methodological Considerations} \label{considerations}

We do not intend to advocate that pilot designs or pilot matching approaches should be used in every scenario.  In the simulations above, the sample size is large and there is an abundance of control individuals which overlap fairly well with the treated population. In such a situation, there are many useful control individuals from which to select the pilot set.  With the expansion of passively collected observational data sets, we anticipate that such situations will become increasingly common. However, in cases where the sample size is small or the overlap of treated and control individuals is sparse, a researcher should consider whether the trade-off to create a pilot set is worthwhile. In this section, we discuss four design considerations which are important to the decision about whether to employ prognostic pilot matching or related pilot designs: 
\begin{enumerate}
    \item Correlation of treatment and prognosis (\ref{consideration-rho})
    \item Trade-offs in sample size (\ref{consideration-n})
    \item Trade-offs in match quality (\ref{consideration-quality})
    \item Fitting the propensity and prognostic models (\ref{consideration-fitting})
\end{enumerate}

\subsubsection{Correlation of Treatment and Prognosis}\label{consideration-rho}

A first consideration is the correlation, $\rho$, between the variation, $\phi(X_i)$, important for determining the treatment assignment, and the variation, $\Psi(X_i)$, important for determining the outcome in the absence of treatment. When the correlation is close to one, all methods unsurprisingly reach their greatest levels of bias, (Figure 2A) because the treatment assignment is tightly correlated with the potential outcomes (i.e. due to the strong observed confounder $X_{i1}$).

It is somewhat counter-intuitive that propensity score matching estimates reach their lowest variance and greatest power in sensitivity analyses when confounding is at its worst (Figures 2B and 3B). This is in part because, when propensity and prognosis are highly correlated, the covariates most important for treatment assignment are also important for prognosis. In such cases, the the optimal propensity score matches often happen to be similar in their potential outcomes. This imposes prognostic similarity between matched observations, reducing the variance (Figure 2B) and increasing robustness to unmeasured confounding (Figure 3B).  

In contrast, the variance of propensity score matching is greatest when treatment assignment is less closely correlated with potential outcome (Figure 2B), since ideal propensity score matches may be quite distant in prognostic score in these situations (Figure 1B) \footnote{In fact when $\rho = 0$, propensity score matching essentially models and adjusts for an instrumental variable ($X_1$), which may actually \textit{increase} inconsistency in estimation \cite{bhattacharya2007instrumental}}. These are the scenarios in which prognostic pilot matching is most protective against variance in estimation, since it explicitly optimizes for within-set prognostic similarity as well as propensity score similarity. This point may apply more generally beyond matching studies alone: the benefit of adjusting for prognosis in addition to propensity for treatment may be greatest when a prognostic score would capture a substantial amount of variation in the data which the propensity score does not.

\subsubsection{Trade-offs in Sample Size}\label{consideration-n}

The most obvious trade-off implicit in any pilot design is that allocating a pilot set decreases the size of the data set used in analysis.  In general, increasing the sample size of a study has the primary goal of reducing the variance in estimation (e.g. increasing $k$ in $1:k$ matching).  A second benefit of a larger sample size is that it may include more high-quality data (e.g. controls which are very closely matched to treated observations). We discuss the former point here, and the latter in section \ref{consideration-quality}.

Roughly speaking, the square root law implies that the standard error of the sample mean tends to decrease at the rate $\frac{1}{\sqrt{n}}$, where $n$ is the sample size. This means the if the sample size is already quite large, aggressively reserving observations for the analysis phase of a study may yield only a vanishingly small increase in precision. In contrast, when the control reserve is very large, it may in fact be variance-\textit{reducing} to ``sacrifice'' some individuals from the analysis set so that the overall matched sets compared in the analysis phase have better prognostic similarity (Figure 2B). This is illustrated in Figure 5. Since full matching uses all of the treated and control individuals, the prognostic pilot full match has approximately 100 fewer control individuals to use for estimation than the alternative methods. However, in these simulations, the variance of the estimator from pilot full matching is actually slightly smaller than from the other methods (Figure 5).

Another aspect of sample size is too often overlooked: More data is not always better when the quality of that data is poor. In fixed-ratio matching, for example, matching more controls ($k$) to each treated individual decreases variance at the cost of \textit{increasing} bias (Figure 2A). Larger values of $k$ can actually give \textit{worse} MSE in many situations by allowing poorer quality matches into the analysis (Supplementary Figures 4 and 7). Some observations are better left discarded entirely than forced into an analysis for which they are unsuitable.

\subsubsection{Trade-offs in Match Quality}\label{consideration-quality}

In some observational data sets, there may be an abundance of ``low-quality'' controls which are very dissimilar from the treatment group, but a paucity of ``high-quality'' controls which are readily comparable to the treated observations. In all pilot designs, the allocation of high quality observations is an important consideration. For example, high-quality controls may be ideal for fitting an accurate model to estimate prognostic scores in the treatment group. However, these controls might also be ideal would-be matches in the final analysis.

For many data sets and study designs, this consideration of \textit{quality} may be much more important than concerns about raw \textit{numbers} of observations. Allocating high-quality controls to the pilot set to fit an accurate prognostic model can force the subsequent matching efforts to compromise on lower quality matches.  This can increase bias and, in some cases, variance (Supplementary Figures 5-7). This effect is most pronounced when (A) $1:k$ matching is performed and $k$ is large, (B) the control reserve is small, or (C) overlap of the treated and control individuals is poor (Figure 2A, Supplementary Figures 4-7). When $k$ is large, bias for all methods is worse because lower quality matches are being carried over to the analysis phase. If the control reserve is small and/or there is little overlap between the treated and control individuals, every control individual which is close in covariate space to the treated individuals is a precious potential match, and sacrificing these individuals means that there are fewer good alternatives to choose from. 

\begin{figure}[h]
\centering
    \includegraphics[scale = 0.8]{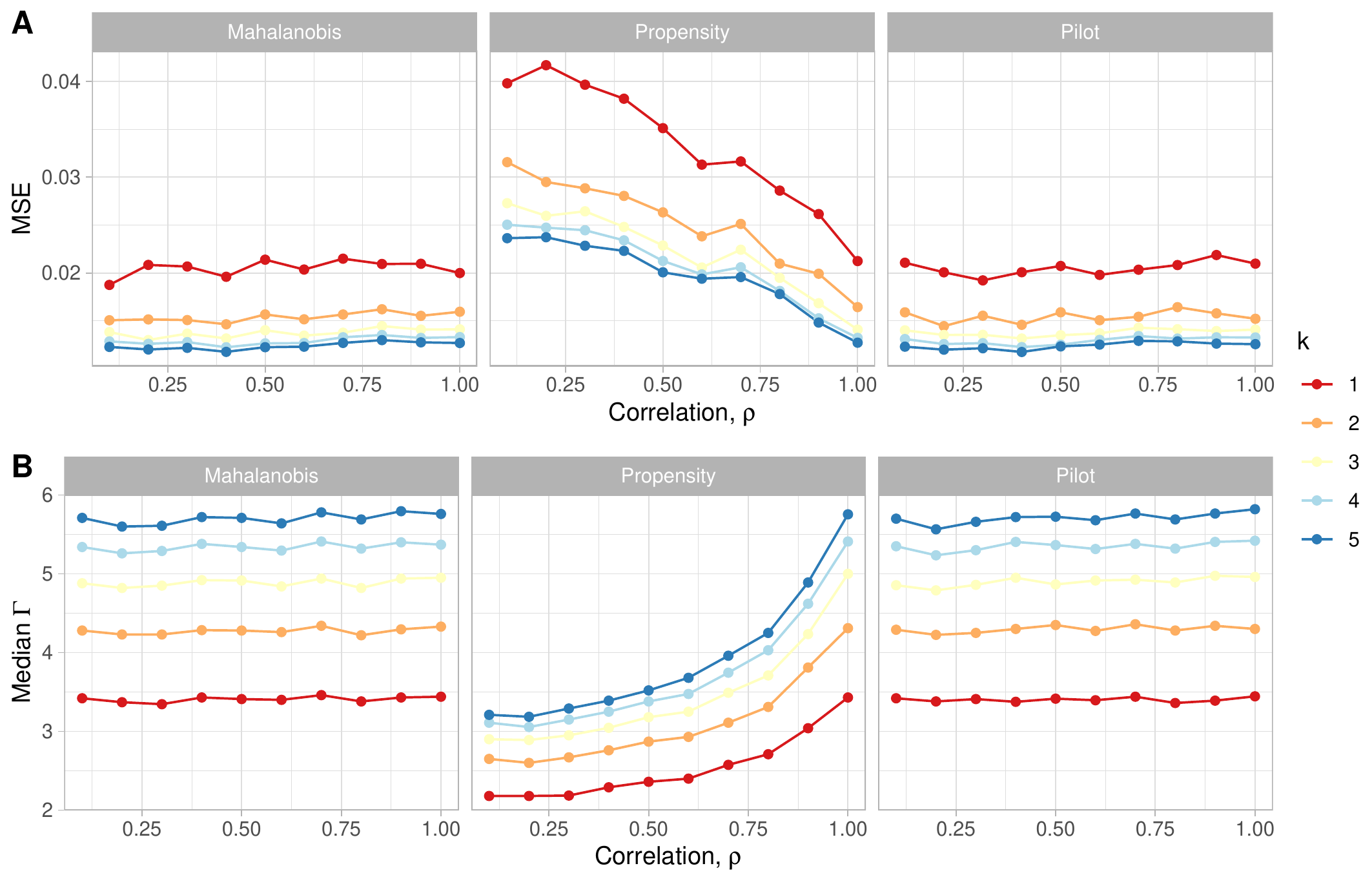}
    \caption{Mahalanobis distance, propensity, and prognostic pilot matching performance in terms of MSE (A) and the median level of confounding, $\Gamma$ at which the study conclusions would no longer be statistically significant (B), in a scenario which supposes that the researcher already knows \textit{a priori} which covariates are most important to treatment assignment and outcome.  All simulation parameters are the same as described in Section 4.2, except that the propensity and prognostic models are both correctly specified, and only $X_{i1}$ and $X_{i2}$ are used in Mahalanobis distance matching.}
\end{figure}

There are a variety of design decisions that can be made to protect against this issue.  First, when overlap is poor -- regardless of whether a pilot design is used -- researchers should consider using lower values of $k$ in fixed-ratio matching, since higher $k$ will force compromises on lower quality matches (Supplementary Figures 4 and 7). One alternative approach to keep bias low while maximizing data usage is to use a matching method that allows for variable ratios of treatment to control individuals across subclasses (see section \ref{fullmatching}). Second, the choice to use a pilot design and the method to select the held-out set should depend on the data: a pilot matching approach is likely to be most appropriate when the control reserve is plentiful, particularly in the region of overlap between treated and control individuals. Further research on clever allocation of the pilot set may make pilot designs useful in a wider variety of scenarios (see section \ref{discussion}).

\subsubsection{Fitting the Propensity and Prognostic Models}\label{consideration-fitting}

A final consideration -- specific to designs which model propensity and prognosis -- the estimation of the score models themselves. While Figure 1 shows the idealized matches that would be selected if the propensity and prognostic scores were precisely known, Figure 7 displays the matches which might be selected in practice based on estimated score models. Importantly, \textit{both} propensity and prognostic pilot matching rely on effective modeling of propensity and/or prognostic scores -- in cases where the model is imperfect, imperfect matches will be selected. Unsurprisingly, we find that the relative performance of prognostic pilot matching is diminished in simulations in which the prognostic model is harder to fit (Supplementary Figure 8). In cases like these, it may be useful to budget more data to the pilot set (if able), or to use more sophisticated modeling techniques (see Supplementary Figure 1 for simulation results where the score models are fit using a lasso regularization.)

In this paper, we primarily consider a situation in which the researcher has no prior information about which covariates are most important to the problem. Thus, in the simulations above, the propensity and prognostic scores were fit on all measured covariates (i.e. both models were over-specified). In this scenario, the propensity and prognostic models act as entirely agnostic methods of identifying the variation most important to the problem. In practice, if the researcher has a prior notion about which covariates are important to the outcome and the treatment assignment, they should consider more precisely specified models. In simulations in which the prognostic and propensity scores are correctly specified, we observe similar if not increased benefit from using the prognostic pilot match compared to the propensity score alone (Figure 6).  Interestingly, when the important covariates are known \textit{a priori}, we find that propensity score achieves lower performance than both competing methods due to larger variances.  This is consistent with the argument by King and Nielson \cite{king2016propensity} that propensity score methods can be variance-\textit{increasing} because of a failure to account for variation unrelated to treatment assignment.

The researcher may also have intuition \textit{a priori} about whether the propensity or the prognostic score will be harder to fit. In some cases, the available measured covariates may make accurate propensity score modeling more difficult than prognostic score modeling, or vice-versa. In some instances when the outcome is continuous and the treatment is binary, the continuous number may be intrinsically easier to model than the binary one. Especially in light of theorem 2, including a prognostic score is likely to be most beneficial when the treatment assignment is difficult to model yet the outcome under the absence of treatment is fairly straightforward.

\begin{figure}[h]
\centering
    \includegraphics[scale = 0.8]{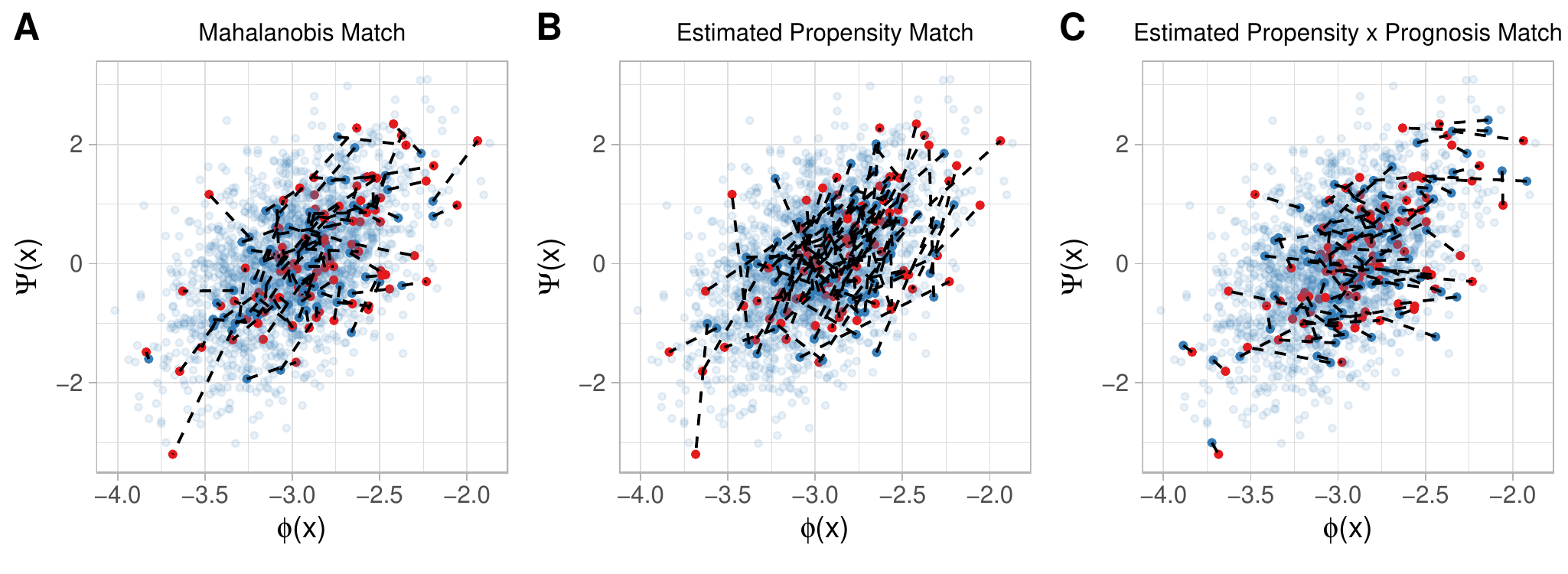}
    \caption{AC plots of empirical 1:1 matches based on Mahalanobis distance in the whole covariate space (A), estimated propensity score (B), and prognostic pilot matching (C). Blue dots represent control individuals, red dots represent treated individuals, and dotted lines connect matched pairs. Data was simulated according to the set up in section \ref{setup} with $\rho = 0.5$.}
\end{figure}

\section{Conclusions}\label{discussion}

Reducing the number of observations used in inference seems a peculiar design choice. But the primary intuition is straightforward: use resources to gain information on how to plan a stronger study design. A fundamental insight regarding the usefulness of pilot designs is that, in general, the standard error in an estimate tends to decrease with $\frac{1}{\sqrt{n}}$, where $n$ is the sample size. Modern observational studies benefit from having access to sample sizes that are unparalleled by randomized controlled trials. However, this means that each additional individual in such an observational study may contribute only minutely in increasing the precision of estimation.  The question then becomes: is there a more thoughtful way to use these data resources in order to improve this study?

The pilot design described here aims to emulate a two-phase, prospective randomized trial, in which a pilot sample is used to identify prognostically important variation, which is then addressed in the design of the main study. In the experimental setting, these design choices reduce the heterogeneity between compared treatment and control subjects, increasing precision and power. However, in the observational context, minimizing the heterogeneity in the prognostic score within matched sets yields an additional benefit. As demonstrated by Rosenbaum \citep{rosenbaum2005heterogeneity} and underscored here, observational studies which compare observations that are more similar in their prognostic scores tend to be more resistant to having their conclusions explained away by unobserved confounding variation. Moreover, as investigated by Leacy and Stuart\cite{leacy2014joint} and Antonelli et al.\cite{antonelli2018doubly} and reiterated here, matching jointly on the prognostic and propensity scores yields study designs which tend to be doubly robust.  Said more informally: a strong prognostic model can make up for errors from a weak propensity model, and vice versa.

Similar to the propensity score, the prognostic score has a variety of promising applications not limited to the matching application here (see, for example \cite{stuart2013prognostic}), which we hope will continue to be explored in future works. Indeed, prognostic score stratification has been employed in the experimental literature, and similar concerns of prognostic score over-fitting have been discussed in this setting \cite{abadie2018endogenous}. A related prognostic score pilot design is implemented by the CRAN package ``strata\_match.''  Researchers looking for more experience and direction on the implementation of a pilot design approach with real or simulated data are encouraged to consider this package and the accompanying documentation \cite{aikens2020stratified}.

From this investigation of prognostic score approaches, the Assignment-Control (AC) plot has emerged as a helpful tool for understanding the underlying structure of a data set and a causal inference problem. In modern observational studies, both the number of observations and the number of covariates tend to be large.  AC plots suggest a dimensionality reduction that can be used to visualize overlap between treated and control individuals in terms of two important aspects of variation: prognosis and likelihood of treatment. This can help a researcher identify issues in the distributions of treated and control individuals which could not be identified by considering propensity alone (e.g. are only very sick patients being treated?). A small, but useful, extension to the AC plot is to include additional dimensions that describe (i) the conditionally independent variation in the encouragement to treatment (e.g., a measurement of the instrumental variable assignment) when considering the formation of matched sets, and (ii) the estimated outcome for individual units under treatment \cite{wijayatunga2018probabilistic, yang2020multiply} -- which may be particularly helpful in settings with heterogeneous treatment response.

Though we made the choice to motivate this paper in the large control reserve setting, in the simulations it becomes apparent that a pilot matching approach may also be quite useful when the number of observations is large but the treatment to control ratio is closer to 1.  Wijayatunga  \cite{wijayatunga2018probabilistic} and Yang et al \cite{yang2020multiply} address an interesting generalization of the prognostic score which models both the potential outcome under the control assignment and the potential outcome of the treatment assignment. While they do not thoroughly address how such score models should be fit, it is likely that a pilot design approach may be applicable. In this paper, while we focused the discussion upon the estimation of the sample average treatment effect among the treated, more work remains to be done articulating the implications of subsetting the sample to fit the pilot set.  A possible extension of the pilot design is to use a cross-validation based approach, in which the data set is repeatedly split into a model-fitting and estimation set.  This is closely related to the framework suggested by Abadie et al. \cite{abadie2018endogenous}, and may be especially useful when the study objective is to estimate a more generalizable average treatment effect (ATE) rather than the sample average treatment effect among the treated (SATT).  It is important to note that matching methods which combine prognostic and propensity score information, as stressed in section \ref{gamma_sensitivity}, improving power in gamma sensitivity analyses may do nothing to change the underlying presence or absence of an unobserved confounder or a treatment effect.  Future work may further develop a framework to address the apparent contradiction illustrated in Figure 4: designs favored based on power in gamma sensitivity analyses can be equivalently biased when an unobserved confounder is at play.

In section \ref{considerations}, this study explored four trade-offs faced by the researcher in designing an observational study using a prognostic pilot matching design.  Some of these trade-offs are more generally important for the selection of any pilot design (i.e. sample size and quality).  In each case, we avoid giving any hard universal cut-offs to define what samples are appropriate for a pilot design.  Since there are many inter-related factors at play, the researcher needs to weigh the different aspects of the problem in order to make the call. Some avenues for future work may center around innovations which make these trade-offs more favorable.  Further advances in the thoughtful selection of the pilot set and clever modeling of the propensity and prognostic scores may make pilot design approaches both more effective and more adaptable to a wider variety of data sets.

\section*{Acknowledgments}
The authors would like to thank Jonathan H. Chen for contributing his computational resources to this project. We would also like to acknowledge contributions of our anonymous reviewers, who suggested many improvements and future directions to this work (for example the suggestion of a resampling approach to estimating the prognostic score). RCA is supported by funding from the National Institutes of Health (T32 LM012409) and a Stanford Graduate Fellowship in Science and Engineering. DG is supported internally by the Stanford Department of Statistics.

\subsection*{Author contributions}

All authors contributed to the design and set-up of the simulations and the development of the methodologies. RCA and DG developed the code, ran the simulations, and generated the figures. RCA developed the mathematical results. RCA, DG, and MB wrote the manuscript. All authors contributed to the editing and revision of the manuscript.

\subsection*{Financial disclosure}

None reported.

\subsection*{Conflict of interest}

The authors declare no potential conflict of interests.

\section*{Supporting information}

The following supporting information is available as part of the online article:

\noindent
\textbf{Figure S1.}
{Simulation results for a scenario in which the score models are fit with a lasso regularization.}

\noindent
\textbf{Figure S2.}
{Simulation results for a scenario in which the number of uninformative covariates is increased}

\noindent
\textbf{Figure S3.}
{Simulation results for a scenario in which there is an unmeasured confounder.}

\noindent
\textbf{Figure S4.}
{Simulation results for a scenario in which there is a smaller control reserve.}

\noindent
\textbf{Figure S5.}
{Example AC plots for the simulation scenario in which the overlap between treated and control individuals is diminished.}

\noindent
\textbf{Figure S6.}
{Example histogram of logit propensity score for the simulation scenario in which the overlap between treated and control individuals is diminished.}

\noindent
\textbf{Figure S7.}
{Simulation results for a scenario in which the overlap between treated and control individuals is diminished.}

\noindent
\textbf{Figure S8.}
{Simulation results for a scenario in which the random noise contributing to the outcome is increased.}

\noindent
\textbf{Figure S9.}
{Simulation results for a scenario in which the treatment effect is heterogeneous.}

\noindent
\textbf{Mathematics Supplement.}
{Additional proofs, notes, and regulatory conditions for the mathematical results.}

\nocite{*}
\bibstyle{AMA}
\bibliography{citations}%

\clearpage

\end{document}


\maketitle

\section{Theorem 1: Properties of a simple matching estimator in terms of the prognostic score}

Suppose we observe $n_T$ treated individuals (indexed by $i$), and suppose each treated individual is matched to exactly one control individual (using any matching scheme). Let $j_{(i)}$ represent the index of the control individual matched to the $i$-th treated individual.  Finally, suppose we additionally assume the following:
\begin{enumerate}
    \item The outcome is continuous, and for all individuals, $Y(0) = \Psi(X) + \epsilon,$ where $\epsilon \iidsim N(0, \sigma^2)$
    \item $\epsilon$ is independent of $\Psi(X)$
    \item The differences in outcomes between matched pairs, $D_i := Y_i - Y_{j_{(i)}}$, are independent of each other.
\end{enumerate}

\begin{theorem}\label{thm1}
Assume the statements above and suppose $\tau$ is a constant additive treatment effect. Let $\Psi(X_i) - \Psi(X_{j_{(i)}})$ denote the difference in prognostic score for a randomly selected treated individual and its matched control.  If we estimate the ATT via $\hat{\tau} := \frac{1}{n_T}\sum_{i = 1}^{n_T} D_i$, then 
\begin{equation}
    Var(\hat{\tau}) = \frac{4\sigma^2}{n_T} +\frac{1}{n_T} Var\left(\Psi(X_i) - \Psi(X_{j_{(i)}})\right),
\end{equation}

\begin{equation}
    Bias(\hat{\tau}) =  E\left[\Psi(X_i) - \Psi(X_{j_{(i)}})\right],
\end{equation}

and
\begin{align}
    MSE(\hat{\tau}) &= \frac{4\sigma^2}{n_T} +\frac{1}{n_T} Var\left(\Psi(X_i) - \Psi(X_{j_{(i)}})\right) +\left(E\left[\Psi(X_i) - \Psi(X_{j_{(i)}})\right]\right)^2 \\
    &\leq \frac{4\sigma^2}{n_T} + E\left[\left(\Psi(X_i) - \Psi(X_{j_{(i)}})\right)^2\right]
\end{align}
Moreover, if assumptions 2 and 3 do not hold, the right hand side of (1) and (3) become upper bounds, and (2) and (4) still hold.

\end{theorem}

\begin{proof}
We will first derive the result for variance.

\begin{align*}
    Var(\hat{\tau}) &= \frac{1}{M}Var(D_i) \\
    &= \frac{1}{M}Var(Y_i - Y_{j_{(i)}}) \\
    &= \frac{1}{M}Var(\tau + Y_i(0) - Y_{j_{(i)}}(0)) \\
    &= \frac{1}{M}Var(\Psi(X_i) + \epsilon_i - \Psi(X_{j_{(i)}}) - \epsilon_{j_{(i)}}) \\
    &= \frac{4\sigma^2}{M} +\frac{1}{M} Var(\Psi(X_i) - \Psi(X_{j_{(i)}})
\end{align*}

If assumption 2 is false then the statement becomes an inequality at line 5.  If assumption 3 is false then the statement becomes an inequality at lines 1-5.

Next, bias:

\begin{align*}
    Bias(\hat{\tau}) &= E\left[D_i - \tau\right] \\
    &= E\left[Y_i(0) + \tau - Y_{j_{(i)}}(0) - \tau \right] \\
    &= E\left[Y_i(0) - Y_{j_{(i)}}(0) \right] \\
    &= E\left[\Psi(X_i) + \epsilon_i - \Psi(X_{j_{(i)}}) - \epsilon_{j_{(i)}}\right] \\
    &= E\left[\Psi(X_i) - \Psi(X_{j_{(i)}})\right]
\end{align*}

Thus, for the MSE we have

\begin{align*}
    MSE(\hat{\tau}) &= \frac{4\sigma^2}{M} +\frac{1}{M} Var(\Psi(X_i) - \Psi(X_{j_{(i)}}) +\left(E\left[\Psi(X_i) - \Psi(X_{j_{(i)}})\right]\right)^2 \\
    &\leq \frac{4\sigma^2}{M} + E\left[\left(\Psi(X_i) - \Psi(X_{j_i})\right)^2\right]
\end{align*}

\end{proof}

\section{Theorem 2: Doubly Robust Estimation}

The following result is a generalized restatement of Theorem 2 in \cite{antonelli2018doubly}, which centers around a doubly robust matching estimator (DRME), in which the models for $\hat{\phi}$ and $\hat{\Psi}$ are estimated using a lasso.  The following is almost immediate from Antonelli et al\cite{antonelli2018doubly}; however, in this form of the theorem, we clarify how the rate of convergence for the estimator from the pilot design depends on: (1) the sample size of the pilot dataset (2) the sample size of the analysis dataset, and (3) the specific method selected to estimate $\hat{\theta}$.

Let $n_{\mathcal{D}'}$ and $n_\mathcal{P}$ be the sample sizes of the analysis and the pilot data sets, respectively. Following the notation of \cite{antonelli2018doubly}, let $\hat{\theta}$ represent a vector of the estimated parameters for the propensity and prognostic models, and for a given $\hat{\theta}$ let $Z = \left(\hat{\phi}(X), \hat{\Psi}(X)\right)$.  Let $\tilde{\theta}$ represent the probabilistic limit of $\theta$ (Note that this may not be the true model parameters, since the models for $\phi$ and $\psi$ may be mis-specified). In keeping with \cite{abadie2006large} and \cite{antonelli2018doubly}, suppose the $M$ matches for the $i^{th}$ individual with score model parameters $\theta$ are given by

\begin{equation}
   \mathcal{J}_M(i, \theta) = \left\{j = 1, ... , n_{\mathcal{D}'}: T_j = 1- T_i, M \geq \sum_{l:T_l = 1-T_i}I\left(\lVert Z_i - Z_l\lVert < \lVert Z_i - Z_j\lVert\right)\right\},
\end{equation}

Where $I$ is an indicator variable and $\lVert\cdot\lVert$ is Mahalanobis distance.  Note that this is slightly different than the matching scheme used in this simulation study and in many study designs, in that each individual (both treated and control) is matched with the $M$ ``nearest" individuals of the opposite treatment assignment in the analysis set, and individuals may be used more than once.  In practice, many studies use matching without replacement, which we consider here in simulation.

Finally, let $\tau(\theta)$ represent the estimator obtained from the matching given by $\mathcal{J}_M(i, \theta)$:

\begin{equation}
    \tau(\theta) = \frac{1}{N_{\mathcal{D}'}}\sum_{i = 1}^{N_{\mathcal{D}'}}(2T_i - 1)\left(Y_i - \frac{1}{M}\sum_{j \in \mathcal{J}_M(i, \theta)}Y_j\right).
\end{equation}

First, we list some conditions necessary for Theorem 2, and then state the proof, which is a slight modification of Antonelli et al. \cite{antonelli2018doubly}.  Readers particularly interested in the mathematics are encouraged to see Antonelli et al. \cite{antonelli2018doubly} and Abadie and Imbens \cite{abadie2006large}.

\subsection*{Conditions necessary for Theorem 2\label{thm_2_assumptions}}

In keeping with Antonelli et al.\cite{antonelli2018doubly}, we define the \textit{matching discrepancy}\footnote{We note that this is a slight departure from the definition of matching discrepancies in Abadie and Imbens \cite{abadie2006large}, which would consider the vector $Z_j(\theta) - Z_i(\theta)$ in $\R^2$, rather than its norm.} between observations $i$ and $j$ given score models parametrized by $\theta$ as $D_{ij(\theta)} = \lVert Z_j(\theta) - Z_i(\theta)\lVert^2$.
Let $f_{D;i;\theta; n_{\mathcal{D}'}}$ represent the probability density function of matching discrepancies between $i$ and observations with the opposite treatment assignment for a given $\theta$ and $n_{\mathcal{D}'}$, and let $D_{i(k)}$ represent the discrepancy between observation $i$ and the $k$-th nearest observation with the opposite assignment.

Following Antonelli et al \cite{antonelli2018doubly}, let $C_{iM} = \frac{D_{i(M)} + D_{i(M+1)}}{2}$, and let $l_{ij}(\theta) = C_{iM} - \lVert Z_j(\theta)  - Z_i(\theta) \lVert^2$.  Last, for any distinct observations $i, j, k$, let $H_{ij}(\theta) = Y_j^2Y_i^2\left(\frac{\partial}{\partial \theta}l_{ij}(\theta)\right)^2\left(\frac{\partial}{\partial \theta}l_{ji}(\theta)\right)^2$ and $H_{ijk}(\theta) = |Y_i||Y_j||Y_k|\left|\frac{\partial}{\partial \theta}l_{ij}(\theta)\right|\left|\frac{\partial}{\partial \theta}l_{ji}(\theta)\right|\left|\frac{\partial}{\partial \theta}l_{ik}(\theta)\right|\left|\frac{\partial}{\partial \theta}l_{ki}(\theta)\right|\left|\frac{\partial}{\partial \theta}l_{kj}(\theta)\right|\left|\frac{\partial}{\partial \theta}l_{jk}(\theta)\right|$. \\

Theorem 2 requires the following:

\begin{enumerate}
    \item The covariates have support, $\mathbb{X}$, which is compact and convex, and the density distribution of the covariates is bounded and bounded away from zero on its support,
    \item SUTVA, strong ignorability, positivity, and no effect modification,
    \item All $(Y_i, T_i, X_i)$ observations in the analysis set are independent draws from the same distribution,
    \item $\int f^2_{D;i;\tilde{\theta}; n_{\mathcal{D}'}}(x)dx < \infty$ and $\int f^2_{D;i;\hat{\theta}n_{\mathcal{D}'}}(x)dx < \infty$ for all $n_{\mathcal{D}'}$ suitably large,
    \item $E\left[H_{ij}(\tilde{\theta})\right] < \infty$ and $E\left[H_{ijk}(\Tilde{\theta})\right] < \infty$ for any $i \neq j \neq k$ and any $n_{\mathcal{D}'}$ suitably large,
    \item Both the propensity and prognostic score models are fit on the pilot set $\mathcal{P}$, and
    \item $E\left[Y(0)|X = x\right]$ and $E\left[Y(1)|X = x\right]$ are Lipschitz functions of $x$ over $\mathbb{X}$.
\end{enumerate}

Note that these are simply the requirements for theorem 2 from Antonelli et al. \cite{antonelli2018doubly}, without those specific to the use of the high-dimensional lasso.  Assumption 6 is necessary to ensure that the matching discrepancies for a fixed $i$ are independent.  In practice, it is likely that the researcher will choose to estimate the prognostic score on $\mathcal{P}$ and the propensity score model on the entire data set $(\mathcal{P}\cup \mathcal{D}')$. Note that a substantial implicit assumption to any doubly robust approach is that there \textit{is} some true score model (propensity, prognosis, or both) which holds for the entire collected data set: treated, and control.  Otherwise the proper specification of one of the score models is a nonsensical claim.

\subsection*{A modified proof for Theorem 2}

\begin{theorem} (Generalization of \cite{antonelli2018doubly})
Suppose that $\hat{\theta}$ is estimated on the pilot data set by some method such that $\lVert\hat{\theta} - \tilde{\theta}\lVert$ converges in probability to 0 at a rate $R(n_\mathcal{P}, p)$. Under regularity conditions described above, and assuming that at least one of the models for propensity or prognosis is correctly specified, then,
\begin{equation}
    \tau - \tau(\hat{\theta}) = O_p\left(n_{\mathcal{D}'}^{-\frac{1}{2}}\right) +  o_p\left(R\left(n_\mathcal{P}, p\right)\right)
\end{equation}

If both $n_\mathcal{P}$ and $n_{\mathcal{D}'}$ are allowed to go to infinity as the sample size increases, and if $R(n_\mathcal{P}, p)$ is bounded as $n_\mathcal{P}, p \rightarrow \infty$, then we have that $\tau(\hat{\theta})$ is a consistent estimator for $\tau$. Moreover, if $p$ is fixed and $R(n_\mathcal{P}, p)$ is bounded as $n_\mathcal{P} \rightarrow \infty$, then $\tau(\hat{\theta})$ is also a consistent estimator for $\tau$.
\end{theorem}

\begin{proof}
This proof essentially follows the proof from \cite{antonelli2018doubly}, with a small generalization. First, let us define the smoothed matching estimator described by \cite{antonelli2018doubly}.  Notice that we can rewrite $\tau(\theta)$ as:

$$\tau(\theta) = \frac{1}{n_{\mathcal{D}'}}\sum_{i = 1}^{n_{\mathcal{D}'}}(2T_i - 1) \left[Y_i - \frac{1}{M} \sum_{j:T_j = 1-T_i}I\left\{C_{iM} - \lVert Z_j(\theta) - Z_i(\theta)\lVert^2 >0 \right\}Y_j\right]$$

\noindent where $C_{iM}$ is as definted in the previous subsection.  

From this, define the smoothed matching estimator,

$$\tau_\Phi(\theta;h_{n_{\mathcal{D}'}}) := \frac{1}{n_{\mathcal{D}'}}\sum_{i = 1}^{n_{\mathcal{D}'}} (2T_i -1) \left[Y_i - \frac{1}{M} \sum_{j:T_j = 1-T_i}\Phi_{h_{n_{\mathcal{D}'}}}(l_{ij}(\theta))Y_j\right],$$

\noindent where $l_{ij}(\theta) = C_{iM} - \lVert Z_j(\theta)  - Z_i(\theta) \lVert^2$ as above, and  $\Phi_{h_{n_{\mathcal{D}'}}}(x) = (1 + e^{-x/h_{n_{\mathcal{D}'}}})^{-1}$. The quantity $h_{n_{\mathcal{D}'}}$ is some appropriately chosen bandwidth of the smoothed estimator, which we will return to momentarily.

Letting $\Tilde{\theta}$ be the probabilistic limit of $\hat{\theta}$, we continue by breaking the error of the matching estimator down as 
\begin{align*}
    \tau(\hat{\theta}) - \tau &= \left[\tau(\hat{\theta})-\tau(\Tilde{\theta})\right] + \left[\tau(\Tilde{\theta}) - \tau\right]\\
    &=\left[\tau(\hat{\theta})-\tau_\Phi(\hat{\theta};h_{n_{\mathcal{D}'}})\right] + \left[\tau_\Phi(\hat{\theta};h_{n_{\mathcal{D}'}})-\tau_\Phi(\Tilde{\theta};h_{n_{\mathcal{D}'}})\right] + \left[\tau_\Phi(\Tilde{\theta};h_{n_{\mathcal{D}'}})-\tau(\Tilde{\theta})\right] + \left[\tau(\Tilde{\theta}) - \tau\right]
\end{align*}

The proof proceeds by showing that for any $Q \geq 0$ and an appropriately chosen bandwidth, $h_{n_{\mathcal{D}'}}$:
\begin{enumerate}
    \item $\tau(\hat{\theta})-\tau_\Phi(\hat{\theta};h_{n_{\mathcal{D}'}}) = o_p\left(N^{-Q}\right)$
    \item $\tau_\Phi(\Tilde{\theta};h_{n_{\mathcal{D}'}})-\tau(\Tilde{\theta}) = o_p\left(N^{-Q}\right)$
    \item $\tau_\Phi(\hat{\theta};h_{n_{\mathcal{D}'}})-\tau_\Phi(\Tilde{\theta};h_{n_{\mathcal{D}'}}) = o_p\left(R_{n_{\mathcal{P}}}, p\right)$
    \item $\tau(\Tilde{\theta}) - \tau = O_p\left(N^{-1/2}\right)$
\end{enumerate}

The proofs for 1,2, and 4 have already been done in \cite{antonelli2018doubly} and \cite{abadie2006large}.  We restate the results here, but for a more thorough derivation it is best to consult those sources.  The only deviation is in the proof for part 3. \\

\textbf{Results 1 and 2: } In the proof of Theorem 2 from \cite{antonelli2018doubly}, the authors show that for an appropriate selection of $h_{n_{\mathcal{D}'}}$ and any $Q\geq0$,

\begin{align*}
    \tau(\hat{\theta})-\tau_\Phi(\hat{\theta};h_{n_{\mathcal{D}'}}) &= o_p\left(n_{\mathcal{D}}^{-Q}\right) \text{, and}\\
    \tau_\Phi(\Tilde{\theta};h_{n_{\mathcal{D}}})-\tau(\Tilde{\theta}) &= o_p\left(n_{\mathcal{D}'}^{-Q}\right).
\end{align*}
\noindent An example of such a bandwidth selection is given by \cite{antonelli2018doubly}.  Moreover, the proofs for these results do not require any specific method for estimating $\hat{\theta}$, and the convergence rate does not depend on $p$.\\

\textbf{Result 4: } Assuming SUTVA, strong ignorability, positivity, and that at least one of the propensity and prognostic score models in correctly specified, then 

$$\tau(\Tilde{\theta}) - \tau = O_p(n_{\mathcal{D}'}^{-\frac{1}{2}}).$$

\noindent This is a direct result of Theorem 1 from \cite{abadie2006large} and Theorem 1 from \cite{antonelli2018doubly}.  A slightly deeper explanation can be found in \cite{antonelli2018doubly}.  Importantly, this rate of convergence depends on the number of variables matched upon but \textit{not} the number of covariates, $p$. Regardless of $p$, the matching is done with respect to 2 continuous features: $\phi(X)$ and $\Psi(X)$ parametrized by $\tilde{\theta}$.\\

\textbf{Result 3:} The only divergence from the proof of Theorem 2 from \cite{antonelli2018doubly} is in the simplification of $\tau_\Phi(\Tilde{\theta};h_{n_{\mathcal{D}'}})-\tau_\Phi(\hat{\theta};h_{n_{\mathcal{D}'}})$. First,  \cite{antonelli2018doubly} shows that 

$$\tau_\Phi(\Tilde{\theta};h_{n_{\mathcal{D}'}})-\tau_\Phi(\hat{\theta};h_{n_{\mathcal{D}'}}) = o_p(1)(\hat{\theta} - \Tilde{\theta}) + O_p\left(\lVert\hat{\theta} - \Tilde{\theta}\lVert^2\right).$$
Again, this fact does not rely on the use of the lasso to estimate $\hat{\theta}$. Rather than assuming that $\hat{\theta}$ is estimated with a lasso, we suppose that $\hat{\theta}$ is estimated by some method so that $\lVert\tilde{\theta} - \hat{\theta}\lVert$ converges in probability to 0 at rate $R({n_\mathcal{P}}, p)$,

$$\tau_\Phi(\Tilde{\theta};h_{n_{\mathcal{D}'}})-\tau_\Phi(\hat{\theta};h_{n_{\mathcal{D}'}}) = o_p\left(R\left(n_\mathcal{P}, p\right)\right).$$

Combining results 1-4 and selecting any $Q \geq \frac{1}{2}$, yields:

$$\tau(\hat{\theta}) - \tau = O_p\left(n_{\mathcal{D}'}^{-\frac{1}{2}}\right) +  o_p\left(R\left(n_\mathcal{P}, p\right)\right)$$

\end{proof}

\bibliographystyle{unsrt}
\bibliography{citations}